\RequirePackage{fix-cm}
\documentclass[smallextended]{svjour3}       
\usepackage{natbib}

\smartqed  
\usepackage{graphicx}
\journalname{Empirical Software Engineering}
\usepackage{amsmath,amssymb,amsfonts}
\usepackage{algorithmic}
\usepackage{textcomp}
\usepackage{rotating}
\usepackage{multirow}
\usepackage[dvipsnames]{xcolor}

\usepackage[nolist]{acronym}
\usepackage{caption,booktabs}
\usepackage[title]{appendix}

\usepackage{hyperref}

\usepackage[dvipsnames]{xcolor}

\usepackage{soul}
\soulregister\cite7
\soulregister\Hl{7}
\soulregister\ref7
\soulregister\pageref7
\LetLtxMacro\origcite\cite

\newcommand{\rev}[1]{#1}
\newcommand{\del}[1]{}

\newcommand{\numberProjects}{54}
\newcommand{\numberMetrics}{14}
\newcommand{\numberCommits}{2,533}
\newcommand{\numberCommitsAll}{125,482}

\begin{document}

\begin{acronym}
\acro{NLP}{Natural Language Processing}
\acro{MLM}{Masked Language Model}
\acro{NSP}{Next Sentence Prediction}
\acro{LLOC}{Logical Lines of Code}
\acro{ITS}{Issue Tracking System}
\acro{VCS}{Version Control System}
\acro{AST}{Abstract Syntax Tree}
\acro{ASTs}{Abstract Syntax Trees}
\acro{DAG}{Directed Acyclic Graph}
\acro{ASAT}[ASAT]{Automated Static Analysis Tool}
\acrodefplural{ASAT}[ASATs]{Automated Static Analysis Tools}
\end{acronym}


\title{What really changes when developers intend to improve their source code:}
\subtitle{A commit-level study of static metric value and static analysis warning changes}

\titlerunning{What really changes when developers intend to improve their source code}      

\author{Alexander Trautsch \and
        Johannes Erbel \and
        Steffen Herbold \and
        Jens Grabowski
}

\institute{Alexander Trautsch\\Institute of Software and Systems Engineering, TU Clausthal, Germany\\
           \email{alexander.trautsch@tu-clausthal.de}
           \vspace{5pt}\\
           Johannes Erbel\\Institute of Computer Science, University of Goettingen, Germany\\
           \email{johannes.erbel@cs.uni-goettingen.de}
           \vspace{5pt}\\
           Steffen Herbold\\Institute of Software and Systems Engineering, TU Clausthal, Germany\\
           \email{steffen.herbold@tu-clausthal.de}
           \vspace{5pt}\\
           Jens Grabowski\\Institute of Computer Science, University of Goettingen, Germany\\
           \email{grabowski@cs.uni-goettingen.de}
}

\date{Received: date / Accepted: date}

\maketitle

\begin{abstract}
Many software metrics are designed to measure aspects that are believed to be related to software quality.
Static software metrics, e.g., size, complexity and coupling are used in defect prediction research as well as software quality models to evaluate software quality.
Static analysis tools also include boundary values for complexity and size that generate warnings for developers.
While this indicates a relationship between quality and software metrics, the exten\del{d}\rev{t} of it is not well understood.
Moreover, recent studies found that complexity metrics may be unreliable indicators for understandability of the source code.
To explore this relationship, we leverage the intent of developers about what constitutes a quality improvement in their own code base.
We manually classify a randomized sample of \numberCommits{} commits from \numberProjects{} Java open source projects as quality improving depending on the intent of the developer by inspecting the commit message.
We distinguish between perfective and corrective maintenance via predefined guidelines and use this data as ground truth for the fine-tuning of a state-of-the art deep learning model for natural language processing.
The benchmark we provide with our ground truth indicates that the deep learning model can be confidently used for commit intent classification. We use the model to increase our data set to \numberCommitsAll{} commits.
Based on the resulting data set, we investigate the differences in size and \numberMetrics{} static source code metrics between changes that increase quality\rev{, as indicated by the developer,} and other changes.
In addition, we investigate which files are targets of quality improvements.
We find that quality improving commits are smaller than other commits. Perfective changes have a positive impact on static source code metrics while corrective changes do tend to add complexity.
Furthermore, we find that files which are the target of perfective maintenance already have a lower median complexity than other files.
Our study results provide empirical evidence for which static source code metrics capture quality improvement from the developers point of view. This has implications for program understanding as well as code smell detection and recommender systems.
\end{abstract}

\section{Introduction}
Software quality is notoriously hard to measure~\citep{Kitchenham1996}. The main reason is that quality is subjective and that it consists of multiple factors.
This idea was formalized by Boehm and McCall in the 70s~\citep{Boehm1976,McCall1977}. Both introduced a layered approach where software quality consists of multiple factors.
The standard ISO \cite{ISO9126} and successor \cite{ISO25010} also approach software quality in this fashion.

All these ideas contain abstract quality factors. However, the question remains what concrete measurements can we perform to evaluate the abstract factors of which software quality consists, i.e., how do we measure software quality.
Some software quality models recommend concrete measurements, e.g., ColumbusQM~\citep{Bakota2011} and Quamoco~\citep{Wagner2012}.
Defect prediction researchers also try to build (machine learning) models to find a function that can map measurable metrics to the number of defects in the source code.
This can also be thought of as software quality evaluation, that tries to map internal software quality, measured by code or process metrics, to external software quality measured by defects~\citep{fenton}.
\rev{The internal and external quality categories can also be mapped to perfective and corrective maintenance categories after \cite{Swanson1976}.
Perfective maintenance should increase internal quality while corrective maintenance should increase external quality. Both categories should increase the overall quality of the software.
To ease the readability, we adopt the perfective and corrective terms defined by Swanson for the rest of the paper when referring to the categories.
For general assumptions, we adopt the internal and external quality terms. Internal quality represents what the developer sees, e.g., structure, size, and complexity while external quality what the user sees, e.g., defects.}

Software quality models and defect prediction models use static source code metrics as a proxy for quality~\citep{Hosseini2017}. The intuition is that complex code, as measured by static source code metrics, is harder to reason about and, therefore, is more prone to errors.
However, recent research by \cite{fmri} showed that measured code complexity is perceived very differently between developers and does not translate well to code understanding.
A similar result was found by \cite{Scalabrino2021} although their work is focused on readability measured in a static way.
Both studies, due to their nature, observe developers in a controlled experiment with code snippets.
To supplement these results, it would be interesting to measure what developers change in their code ``in the wild'' to improve software quality\del{.}
\rev{and if their intent matches what we can measure, e.g., if complexity is reduced in a change that intents to improve quality.}

While there are multiple publications on maintenance or change classification after \cite{Swanson1976}, e.g.,~\cite{Mockus2000}, \cite{Mauczka2012}, \cite{Levin2017} and \cite{Honel2019}, we are not aware of publications that investigate differences between multiple software metrics for corrective and perfective maintenance as well as other changes. The inclusion of other changes results in computational effort as we need every metric for every file in every commit. However, we are able to
provide this data via the SmartSHARK ecosystem~\citep{Trautsch2017,Trautsch2020}. This additional effort allows us to infer if categories of changes are different when regarding all changes of a software project.
Most recent work focuses on certain aspects instead of a generic overview, e.g., how software metric values change when code smells are removed~\citep{Bavota2015} or refactorings are applied~\citep{Bavota2015,alshayeb2009,Pantiuchina2020}.
However, we believe that taking a step back from focused approaches and investigating generic quality improvements is worthwhile.
A generic overview has the advantage of mitigating possible problems that can occur for narrow meaning keywords of topically focused approaches while at the same time providing a cohesive overview.
Moreover, it allows for generic statements about software quality evolution based on this information and can complement focused approaches. 

In this work, we find changes that increase the quality, while we measure current, previous and delta of common source code metric values used in a current version~\citep{Bakota2014} of the Columbus quality model~\citep{Bakota2011}.
We use the commit message contained in each change to find commits where the intent of the developer is to improve software quality.
\del{This provides us with a view of corrective and perfective maintenance commits if we think of this approach with the classification by Swanson. Perfective maintenance should increase internal quality while corrective maintenance should increase external quality. Both categories should increase the overall quality of the software. To ease the readability, we adopt the perfective and corrective terms defined by Swanson for the rest of the paper when referring to the categories. For general assumptions, we adopt the internal and external quality terms. Internal quality represents what the developer sees, e.g., structure, size, and complexity while external quality what the user sees, e.g., defects.}
\rev{This provides us with a view of corrective and perfective maintenance commits.}

Within our study, we first classify the commit intent for a sample of \numberCommits{} commits from \numberProjects{} open source projects manually.
The manual classification is provided by two researchers according to predefined guidelines.
According to the overview of previous research in this area provided by \cite{Alomar2020} our study would be the largest manual classification study of commits.
We use this data as ground truth to fine-tune a state-of-the-art deep learning model for natural language processing that was pre-trained exclusively on software engineering data~\citep{seBERT}.
After we determine the performance of the model, we classify all commits, increasing our data to \numberCommitsAll{} commits.

We use the automatically classified data to conduct a two part study. \rev{The} first part is a confirmatory study into the expected behavior of metric values for quality increasing changes.
Expected behaviour, e.g., complexity is reduced in quality increasing changes is derived as hypothesis from existing quality models and the related literature.

In case our data matches the expected behavior from the literature, we can confirm the postulated theories and provide evidence in favor of using the measurements.
Otherwise, we try to establish which metrics may be unsuitable for quality estimation, including the potential reasons.
Even further, we determine whether metrics used in software quality models are impacted by quality increasing maintenance, therefore providing an evaluation for software quality measurement metrics.

The second part of our study is of exploratory nature.
We investigate which files are the target of quality improvements by the developers. We explore whether only complex files are receiving perfective changes and which metric values are indicative of corrective changes.
This provides us with data for practitioners and static analysis tool vendors for boundary values which are likely to have a positive impact on the quality of source code from the perspective of the developers.

\noindent
Overall, our work provides the following contributions:
\begin{itemize}
    \item A large data set of manual classifications of commit intents with improving internal and external quality categories.
    \item A confirmatory study of size and complexity metric value changes for quality improvements.
    \item An exploratory study of size and complexity metric values of files that are the target of quality improvements.
    \item A fine-tuned state-of-the-art deep learning model for automatic classification of commit intents.
\end{itemize}

\noindent
The main findings of our study are the following:
\begin{itemize}
    \item We confirm previous work that quality increasing commits are smaller than other changes.
    \item While perfective changes have a positive impact on most static source code metric values, corrective changes have a negative impact on size and complexity.
    \item The files that are the target of perfective changes are already less complex and smaller than other files.
    \item The files that are the target of corrective changes are more complex and larger than other files.
\end{itemize}
The remainder of this paper is structured as follows.
In Section~\ref{sec:research_questions}, we define our research questions and hypotheses. In Section~\ref{sec:related_work}, we discuss the previous work related to our study.
Section~\ref{sec:case_study} contains our case study design with descriptions for subject selection as well as data sources and analysis procedure.
In Section~\ref{sec:results}, we present the results of our case study and discuss them in Section~\ref{sec:discussion}.
Section~\ref{sec:threats_to_validity} lists our identified threats to validity and Section~\ref{sec:conclusion} closes with a conclusion of our work.

\section{Research Questions and Hypotheses}\label{sec:research_questions}

In our study, we answer two research questions.
\begin{itemize}
    \item \textbf{RQ1:}~\emph{Does developer intent to improve internal or external quality have a positive impact on software metric values?}\\
Previous work provides us with certain indications about the impact on software metric values. This is part of our confirmatory study, and we derive two hypotheses from previous work
regarding how size and software metric values should change for different types of quality improvement.
We formulate our assumptions as hypothesis and test these in our case study.
\begin{itemize}
    \item\textbf{H1:} \textbf{Intended quality improvements are smaller than other changes.}\\
	\cite{Mockus2000} found that corrective changes modify fewer lines while perfective changes delete more lines.
    \cite{Purushothaman2005} also observed more deletions for perfective maintenance and an overall smaller size of perfective and corrective maintenance.
    \rev{Both studies provide measurements we base our hypothesis on. While they are using the same closed source project we will be able to see if our assumption holds for our multiple Java open source projects.}

    \cite{Honel2019} used size-based metrics as additional features for an automated approach to classify maintenance types. They found that the size-based metric values increased the classification performance.
    Moreover, just-in-time quality assurance~\citep{Kamei2013} builds on the assumption that changes and metrics derived from these changes can predict bug introduction, meaning there should be a difference.
    Therefore, we hypothesize that corrective as well as perfective maintenance consist of smaller changes.
    Addition of features should be larger than both\rev{,} and therefore we assume that the categories we are interested in, perfective and corrective, are smaller than other changes.

    \item\textbf{H2:} \textbf{Intended quality improvements impact software quality metric values in a positive way.}\\
    In this paper, we focus on metrics used in the Columbus Quality Model \citep{Bakota2011,Bakota2014}. The metrics are specifically chosen for a quality model so they should provide different measurements based on their maintenance category.
    Prior research, e.g., \cite{chavez2017} and \cite{Stroggylos2007} found that refactorings, which are part of our classification, have a measurable impact on software metric values.
	We hypothesize that an improvement consciously applied by a developer via a perfective commit has a measurable, positive impact on software metric values.
    Positive means that we expect a value change direction of the metric value, e.g., complexity is reduced. We note our expected direction for each metric together with a description in Table~\ref{tbl:metrics}.

    Defect prediction research assumes a connection between software metrics and external software quality in the form of bugs. 
    While most publications in defect prediction are not investigating the impact of single bug fixing changes the most common datasets all contain coupling, size and complexity metrics as independent variables, e.g., \citep{Jureczko2010,mdp,DAmbros2012}, see also the systematic literature review by \cite{Hosseini2017}.
    We hypothesize that fixing bugs via corrective commits has a measurable, positive impact on software metric values.
    \rev{While a bug fix may add complexity, our study compares bug fix changes with all other changes including feature additions.
    Therefore, we do not hypothesize that bug fixing decreases complexity generally, but that it is decreasing complexity in comparison to all other changes.}
    \rev{In contrast to \textbf{H1} we are not able to compare our results to concrete studies as we are not aware of a study that investigates metric value changes of perfective and corrective changes and compares them against all other changes.}
    \rev{We are instead trying to validate the assumption that quality improvements should have a positive impact on software quality metrics as they are found to improve detection of defects~\citep{ferenc2005}.}
\end{itemize}
\end{itemize}
Our second research question is exploratory in nature.
\begin{itemize}
    \item\textbf{RQ2:} \emph{What kind of files are the target of internal or external quality improvements?}\\
The first part of our study provides us with information about metric value changes for quality increasing commits.
In this part, we are exploring which files are the target of quality increasing commits.
We are interested in how complex, e.g., via cyclomatic complexity, a file is on average that receives perfective maintenance.
Moreover, on the external quality side we are interested in which files are receiving corrective changes.
Due to the exploratory nature of this research question, we do not derive hypotheses.
\end{itemize}

\section{Related Work}
\label{sec:related_work}

We separate the discussion of the related work into publications on the classification of changes and publications on the relation between quality improvements and software metrics.

Most prior work that follows a similar approach to ours is concerned with specific types of quality improving changes, e.g., refactoring and removal of code smells.
We note that some code smell detection is based on internal software quality metrics, which we use in our study.

We first present previous research related to the first phase of our study, i.e., classification of changes with respect to maintenance types.
\cite{Mockus2000} study changes in a large system and identified reasons for changes.
They find that a textual description of the change can be used to identify the type of change with a keyword based approach which they validated with a developer survey.
The authors classified changes to Swansons maintenance types. They find that corrective and perfective changes are smaller and that perfective changes delete more lines than other changes.
\cite{Mauczka2012} present an automatic keyword based approach for classification into Swansons maintenance types.
They evaluate their approach and provide a keyword list for each maintenance type together with a weight.

\cite{Fu2015} present an approach for change classification that uses latent drichtlet allocation.
They study five open source projects and classify changes into Swansons maintenance types together with a \emph{not sure} type.
The keyword list of their study is based on \cite{Mauczka2012}.

\cite{Mauczka2015} collect developer classifications for three different classification schemes.
Their data contains 967 commits from six open source projects.
While the developers themselves are the best source of information, we believe that within the guidelines of our approach our classifications are similar to those of the developers.
We evaluate this assumption in Section~\ref{sec:guidelines}.

\cite{Yan2016} use discriminative topic modeling also based on the keyword list by \cite{Mauczka2012}.
They focus on changes with multiple categories.
\cite{Levin2017} improve maintenance type classification by utilizing source code in addition to keywords.
This is an indication that metric values which are computed from source code are impacted by different maintenance types.

\cite{Honel2019} use size metrics as additional features for automated classification of changes.
In our study, we first classify the change and then look at how this impacts size and spread of the change.
However, the differences we found in our study support the assumption that size-based features can be used to distinguish change categories.

More recently, \cite{Wang2021} also analyze developer intents from the commit messages. They focus on large review effort code changes instead of quality changes or maintenance types.
They also use a keyword based heuristic for the classification. They do not, however, include a perfective maintenance classification.

\cite{Ghadhab2021} also use a deep learning model to classify commits. They use word embeddings from the deep learning model in combination with fine\rev{-}grained code changes to classify into Swansons maintenance categories.
In contrast to Ghadhab et al., we do not include code changes in our automatic classifications and focus on the commit message.

The classification of changes for the ground truth in our study is based on manual inspection by two researchers instead of a keyword list. We specify guidelines for the classification procedure which enable other researchers to replicate our work.
To accept or reject our hypotheses, we only inspect internal and external quality improvements which would correspond to the perfective and corrective maintenance types by Swanson.
In contrast to the previous studies, we relate our classified changes also to a set of static software metrics.

We now present research related to our second phase of our study, the relation between intended quality improvements and software metrics.
\cite{Stroggylos2007} found changes where the developers intended a refactoring via the commit message. The authors then measured several source code metrics to evaluate the quality change.
In contrast to the work of \cite{Stroggylos2007}, we do not focus on refactoring keywords. Instead, we consider refactoring as a part of our classification guidelines.
Moreover, our aim is to investigate whether the metrics most commonly used as internal quality metrics (see also \cite{SLRRefactoringQuality}) are the ones that are changing if developers perform quality improving changes including refactoring.

\cite{Fakhoury2019} investigated the practical impact of software evolution with developer perceived readability improvements on existing readability models.
After finding target commits via commit message filtering they applied state\rev{-}of\rev{-}the\rev{-}art readability models before and after the change and investigated the impact of the change on the resulting readability score.

\cite{Pantiuchina2018} analyze commit messages to extract the intent of the developer to improve certain static source code metrics related to software quality.
In contrast to their work, we are not extracting the intent to improve certain static code metrics but instead focus on overall improvement to measure the delta of a multitude of metrics between the improving commit and its parents.
Developers may not use the terminology Pantiuchina et al.~base their keywords on, e.g., instead of writing reduce coupling or increase cohesion the developer may simply write refactoring or simplify code.

In contrast to the previous studies, we relate developer intents to improve the quality either by perfective maintenance or by corrective maintenance to change size metrics and static source code metrics.
In addition, we also look at mean static source code metrics per file which are the target of quality improvements.

\section{Case Study Design}
\label{sec:case_study}
\del{The goal of our case study is to gather empirical data about changes due to developer intents for improving the quality of the code base.}
\rev{The goal of our case study is to gather empirical data about what changes when a developer intents to improve the quality of the code base in comparison to other changes.}
To achieve this, we first sample a number of commits from our selected study subjects.
This sample is classified by two researchers into two categories of quality improving and other changes.
\rev{The classification into categories is only done via the commit message as it expresses the intent of the developer on what the change should achieve.}

This data is then used to train a model that can confidently classify the rest of our \del{data}\rev{commit messages}.
The classified commits are then used to investigate the static source code metric value changes to accept or reject our hypotheses in the confirmatory part of our study.
After that, we investigate the metric values before the change is applied in the exploratory part of our study.

\subsection{Data and Study Subject Selection}
\begin{table}
	\centering
    \caption{Case study subjects with time frame and distribution of commits. All considered commits (\#C), sample size (\#S), sample perfective commits (\#SP), sample corrective commits (\#SC), all perfective commits (\#AP), all corrective commits (\#AC)}\label{tbl:distribution_projects}
    \begin{tabular}{llrrrrrr}
    \toprule
    Project & Timeframe & \#C & \#S & \#SP & \#SC & \#AP & \#AC\\
    \midrule
archiva & 2005-2018 & 3,914 & 79 & 35 & 17 & 1,478 & 1,005\\
calcite & 2012-2018 & 1,987 & 40 & 8 & 14 & 565 & 665\\
cayenne & 2007-2018 & 3,738 & 75 & 31 & 14 & 1,470 & 1,007\\
commons-bcel & 2001-2019 & 884 & 18 & 9 & 6 & 588 & 171\\
commons-beanutils & 2001-2018 & 577 & 12 & 5 & 2 & 317 & 130\\
commons-codec & 2003-2018 & 828 & 17 & 12 & 1 & 619 & 76\\
commons-collections & 2001-2018 & 1,827 & 37 & 27 & 3 & 1,185 & 200\\
commons-compress & 2003-2018 & 1,598 & 32 & 17 & 6 & 873 & 317\\
commons-configuration & 2003-2018 & 2,075 & 42 & 23 & 7 & 1,027 & 253\\
commons-dbcp & 2001-2019 & 1,034 & 21 & 15 & 3 & 672 & 211\\
commons-digester & 2001-2017 & 1,256 & 26 & 16 & 0 & 744 & 113\\
commons-imaging & 2007-2018 & 682 & 14 & 10 & 2 & 476 & 96\\
commons-io & 2002-2018 & 1,036 & 21 & 15 & 3 & 613 & 171\\
commons-jcs & 2002-2018 & 788 & 16 & 10 & 1 & 400 & 162\\
commons-jexl & 2002-2018 & 1,469 & 30 & 20 & 1 & 873 & 199\\
commons-lang & 2002-2018 & 3,261 & 66 & 50 & 6 & 2,182 & 420\\
commons-math & 2003-2018 & 4,675 & 94 & 66 & 10 & 2,981 & 574\\
commons-net & 2002-2018 & 1,092 & 22 & 13 & 5 & 585 & 246\\
commons-rdf & 2014-2018 & 529 & 11 & 9 & 0 & 341 & 35\\
commons-scxml & 2005-2018 & 479 & 10 & 6 & 2 & 256 & 76\\
commons-validator & 2002-2018 & 1,573 & 32 & 18 & 6 & 900 & 296\\
commons-vfs & 2002-2018 & 1,136 & 23 & 11 & 8 & 628 & 207\\
eagle & 2015-2018 & 582 & 12 & 5 & 4 & 104 & 199\\
falcon & 2011-2018 & 1,547 & 31 & 7 & 13 & 255 & 676\\
flume & 2011-2018 & 1,489 & 30 & 5 & 14 & 266 & 591\\
giraph & 2010-2018 & 854 & 18 & 4 & 6 & 201 & 281\\
gora & 2010-2019 & 569 & 12 & 3 & 4 & 182 & 141\\
helix & 2011-2019 & 2,199 & 44 & 8 & 9 & 552 & 580\\
httpcomponents-client & 2005-2019 & 2,399 & 48 & 22 & 16 & 1,113 & 639\\
httpcomponents-core & 2005-2019 & 2,598 & 52 & 25 & 12 & 1,326 & 544\\
jena & 2002-2019 & 8,698 & 174 & 88 & 34 & 4,163 & 1,424\\
jspwiki & 2001-2018 & 4,326 & 87 & 32 & 25 & 1,523 & 941\\
knox & 2012-2018 & 1,131 & 23 & 3 & 10 & 266 & 306\\
kylin & 2014-2018 & 6,789 & 136 & 40 & 40 & 1,904 & 2,163\\
lens & 2013-2018 & 1,370 & 28 & 9 & 9 & 321 & 479\\
mahout & 2008-2018 & 2,075 & 42 & 16 & 15 & 836 & 467\\
manifoldcf & 2010-2019 & 2,867 & 58 & 10 & 21 & 602 & 1,164\\
mina-sshd & 2008-2019 & 1,281 & 26 & 10 & 6 & 381 & 396\\
nifi & 2014-2018 & 3,299 & 66 & 12 & 18 & 592 & 1,052\\
opennlp & 2008-2018 & 1,763 & 36 & 22 & 6 & 805 & 275\\
parquet-mr & 2012-2018 & 1,228 & 25 & 7 & 9 & 439 & 316\\
pdfbox & 2008-2018 & 8,256 & 166 & 81 & 69 & 3,934 & 2,904\\
phoenix & 2014-2019 & 7,835 & 157 & 23 & 83 & 828 & 4,545\\
ranger & 2014-2018 & 2,213 & 45 & 10 & 20 & 434 & 908\\
roller & 2005-2019 & 2,435 & 49 & 15 & 13 & 869 & 723\\
santuario-java & 2001-2019 & 1,455 & 30 & 14 & 5 & 627 & 406\\
storm & 2011-2018 & 2,839 & 57 & 24 & 9 & 987 & 716\\
streams & 2012-2019 & 911 & 19 & 7 & 2 & 264 & 196\\
struts & 2006-2018 & 2,945 & 59 & 21 & 18 & 1,191 & 682\\
systemml & 2012-2018 & 3,860 & 78 & 21 & 25 & 921 & 1,416\\
tez & 2013-2018 & 2,359 & 48 & 8 & 27 & 443 & 1,223\\
tika & 2007-2018 & 2,581 & 52 & 11 & 10 & 705 & 740\\
wss4j & 2004-2018 & 2,455 & 50 & 22 & 10 & 712 & 702\\
zeppelin & 2013-2018 & 1,836 & 37 & 11 & 6 & 333 & 699\\
\midrule
 & & 125,482 & 2,533 & 1,022 & 685 & 47,852 & 35,124\\
    \bottomrule
    \end{tabular}
\end{table}

The data used in our study is a SmartSHARK~\citep{Trautsch2017} database taken from \cite{Trautsch2020a}.
We use all projects and commits in the database. However, only commits that change production code and which are not empty are considered.
For each change in our data, we extract a list of changed files, the number of changed lines, the number of hunks\footnote{An area within a file that is changed.}, and the delta as well as the previous and current value of source code metrics from the changed files between the parent and the current commit.
To create our ground truth sample, we randomly sample 2\% of commits per project rounded up for manual classification.

The data consists of Java open source projects under the umbrella of the Apache Software Foundation\footnote{https://www.apache.org}.
All projects use an issue tracking system and were still active when the data was collected. Each project consist of at least 100 files and 1000 commits and is at least two years old.
Table~\ref{tbl:distribution_projects} shows every project, the number of commits and the years of data we consider for sampling.
In addition, we include the number of perfective and corrective commits for our ground truth and final classification.

\subsection{Change Type Classification Guidelines}\label{sec:guidelines}

As we are not relying on a keyword based approach and there is no existing guideline for this kind of classification, we created a guideline based on \cite{Herzig2013}.
Our ground truth consists of a sample of changes which we manually classified into perfective, corrective, and other changes.
\rev{We do not consider adaptive changes as separate a category. Instead, we include them in the other changes. The reason is that we focus on internal and external quality improvements and map perfective to internal quality and corrective to external quality.}
Every commit message is inspected independently by two researchers with software development experience. The inspection is using a graphical frontend that loads the sample and displays the commit message which can then be assigned a label by each researcher independently.
If the commit message does not provide enough information, we inspect additional linked information in the form of bug reports or the change itself.
In case of a link between the commit message and the issue tracking system, we inspect the bug report and determine if it is a bug according to the guidelines by \cite{Herzig2013}. We perform this step because the reporter of a bug sometimes assigns a wrong type. We defined the guidelines listed in Table~\ref{tbl:guidelines} used by both researchers for the classification of changes.
The deep learning model for our final classification of intents only receives the commit messages. This is a conscious trade-off. On the one hand we want the ground truth to be as exact as possible, on the other hand we want to keep the automatic intent classification as simple as possible. The results of our fine-tuning evaluation (Table~\ref{tbl:sebert_performance}) show that the model does not need the additional data from changes and issue reports to perform well.

Both researchers achieve a substantial inter-rater agreement~\citep{kappa} with a Kappa score of 0.66~\citep{cohenkappa}.
Disagreements are discussed and assigned a label both researchers agree upon after discussion.
The disagreement front end shows both prior labels anonymized in random order.

\begin{table}
    \caption{Classification rules and examples, footnotes denote different commit messages from our data.}\label{tbl:guidelines}
    \centering
    \begin{tabular}{p{0.96\textwidth}}
        \toprule
        A change is classified as \textit{perfective} if\ldots
        \begin{enumerate}
            \item the commit message says code is removed or marked as deprecated.
            \item code is moved to new packages.
            \item generics are introduced, new Java features are used, existing code is switched to collections, or class members are switched to final.
            \item documentation is improved or example code is updated.
            \item static analysis warnings are fixed even though no related bug is reported.
            \item code is reformatted or the readability is otherwise improved (e.g. whitespace fixes or tabs to spaces).
            \item existing code is cleaned up, simplified, or its efficiency improved.
            \item dependencies are updated.
            \item developer tooling is improved, e.g., build scripts or logging facilities.
            \item the repository layout is cleaned, e.g., by removing compiled code or maintaining .gitignore files.
            \item tests are improved or added.
        \end{enumerate}
        \begin{itshape}
            \textbf{Examples:}
            Eliminated unused private field.  JIRA: DBCP-255\footnote{commons-dbcp:fd59279}
            \textnormal{Because of other null checks it was already impossible to use the field. Thus, this is clean up.}
            [CODEC-127] Non-ascii characters in source files\footnote{commons-codec:096c0cc} \textnormal{While the linked issue is a bug, it only affects IDEs for developers and not the compiled code. Thus, this is an improvement of developer tooling.}
            JEXL-240: Javadoc\footnote{commons-jexl:e3c80ca} \textnormal{The message indicates that this commit only improved the code comments. Therefore, it is classified as perfective.}
        \end{itshape}\\

        \midrule

        A change is classified as \textit{corrective} if\ldots
        \begin{enumerate}
            \item the commit message mentions bug fixes.
            \item the commit message or the linked issue mentions that a wrong behaviour is fixed.
            \item the commit message or the linked issue mentions that a NullPointerException is fixed.
            \item a bug report is linked via the commit message that is of type bug and is not just a feature request in disguise (see \cite{Herzig2013}).
        \end{enumerate}
        \begin{itshape}
            \textbf{Examples:}
            KYLIN-940 ,fix NPE in monitor module ,apply patch from Xiaoyu Wang\footnote{kylin:623585f}
            \textnormal{This fixes a NullPointerException that is visible to the end user.}
            owl syntax checker (bug fixes)\footnote{jena:15ecc3c}
            \textnormal{Fixes a wrong behavior.}
        \end{itshape}\\

        \midrule

        A change is classified as \textit{other} if\ldots
        \begin{enumerate}
            \item the commit message mentions feature or functionality addition.
            \item the commit message mentions license information or copyrights changes.
            \item the commit message mentions repository related information with unclear purpose, e.g., merges of branches without information, tagging of releases.
            \item the commit message mentions that a release is prepared.
            \item an issue is linked via the commit message that requests a feature.
            \item any of the 1-5 are tangled with a perfective or corrective classification.
        \end{enumerate}
        \begin{itshape}
            \textbf{Examples:}
            KYLIN-715 fix license issue\footnote{kylin:d9cf556}
            \textnormal{License changes or additions are not direct improvements of source code.}
            Support the alpha channel for PAM files. Fix the alpha channel order when reading and writing. Add various tests.\footnote{commons-imaging:9c563ec}
            \textnormal{This change adds support for a new feature, fixes something and adds tests, it is therefore highly tangled and we do not classify it as either or both.}
        \end{itshape}\\

        \bottomrule
    \end{tabular}
\end{table}

In contrast to the classification by \cite{Mauczka2015} and \cite{Hattori2008}, we do not categorize release tagging, license or copyright corrections as perfective.
Our rationale is that these changes are not related to the code quality, which is our main interest in this study.

\rev{In \cite{Mauczka2015} the researchers selected six projects and seven developers with personal commitment and provided the developers with the commit messages that they then labeled according to different classification schemes.
One of which is the Swanson classification which matches our study.
Each developer labeled a sample of commit messages from their respective project.
As we are focused on Java we also use the Java projects of the \cite{Mauczka2015} dataset to validate our guidelines.}

\rev{Two authors of this paper re-classified the Java projects from \cite{Mauczka2015}: Deltaspike, Mylyn-reviews and Tapiji.
The commit messages were classified separately first. Disagreements were then resolved together in a separate session.
In the first session both authors achieve a substantial inter-rater agreement~\citep{kappa} with a Kappa score of 0.62~\citep{cohenkappa}.}

\rev{Aside from the classification differences regarding release tagging, license or copyright changes, we noticed further differences.
Several commits contain some variation of ``minor bugfixes'' which are classified as perfective maintenance by the developers or both corrective and perfective, whereas we classify them as corrective.
Additionally, code removal or test additions were not classified as perfective changes by the developers, but rather as corrective changes.
This reveals a difference of perspective between researchers and developers.
We consider pure code removal and test additions as perfective instead of corrective as we think of corrective changes as improving external quality, e.g., by fixing a customer facing bug.
The data also contains clean\rev{-}up and removal messages without a hint of an underlying bug which are classified as corrective by the developers.
Based on the information available to us, we cannot decide if these are misclassifications by the developers, the result of differences in the classification guidelines, or misclassifications by us due to lack of in-depth knowledge about the projects.}

\rev{The authors achieve a substantial inter-rater agreement~\citep{kappa} with the developers yielding a Kappa score of 0.63~\citep{cohenkappa}.}

\subsection{Deep Learning For Commit Intent Classification}

\begin{table}
    \centering
    \caption{Change classification model performance comparison.}\label{tbl:sebert_performance}
    \begin{tabular}{lrrrp{4.5cm}}
        \toprule
        Model & Acc. & F1 & MCC & Description\\
        \midrule
        \cite{seBERT} & 0.80 & 0.79 & 0.70 & BERT model pre-trained on software engineering data, fine-tuned with only commit messages\\
        \cite{Ghadhab2021} & 0.78 & 0.80 & - & BERT model pre-trained on natural language, includes code changes.\\
        \cite{Gharbi2019} & - & 0.46 & - & Multi-label active learning, only commit message\\
        \cite{Levin2017} & 0.76 & - & - & Keywords and code changes, Random Forest model\\
        \cite{Honel2019} & 0.80 & - & - & LogitBoost model, includes code density.\\
        \bottomrule
    \end{tabular}
\end{table}
In order to use all available data, we use a deep learning model that classifies all data which is not manually classified into perfective, corrective or other.
Due to the size of state-of-the-art deep learning models and the computing requirements for training them, a current best practice is to use a pre-trained model which was trained unsupervised on a large data set.
The model is then fine-tuned on labeled data for a specific task.

To achieve a high performance, we use seBERT \citep{seBERT}, a model that is pre-trained on textual software engineering data in two common \ac{NLP} tasks.
\ac{MLM} and \ac{NSP} which predict randomly masked words in a sentence and the next sentence respectively. Combined, this allows the model to learn a contextual understanding of the language.
While \cite{seBERT} include a similar benchmark based on our ground truth data, it only used the perfective label, i.e., a binary classification to demonstrate text classification for software engineering data.
In our study, we measure performance of the multi-class case with all three labels, perfective, corrective and other.
Within this study, we first use our ground truth data to evaluate the multi-class performance of the model.
We perform a 10x10 cross-validation which splits our data into 10 parts and uses 9 for fine-tuning the model and one for evaluating the performance.
The fine-tuning itself splits the data into 80\% training and 20\% validation. The model is then fine-tuned and evaluated on the validation data for each epoch.
At the end the best epoch is chosen to classify the test data of the fold.
This is repeated 10 times for every fold which yields 100 performance measurements.

Our experiment shows sufficient performance comparable to other state-of-the-art models for commit classification.
We provide the final fine-tuned model as well as the fine-tuning code as part of our replication kit for other researchers. Performance wise our model is comparable to~\cite{Ghadhab2021} and improves performance compared other studies, e.g., \cite{Gharbi2019,Levin2017}.
However, we note that we fine-tuned the model with only the labels used in our study, i.e., perfective, corrective and other. Therefore, it cannot be used or directly compared with models that support other commit classification labels. This would require the same data and labels, we can only compare the given model performance metrics, which we do in Table~\ref{tbl:sebert_performance}.
If we look at the overview of commit classification studies by \cite{Alomar2020} we can see that our model outperforms the other models for comparable tasks where accuracy or F-measure is given. While this is evidence that our model can perform our required commit intent classification a throughout comparison of different commit intent classification approaches is not within the scope of this study.

\subsection{Metric Selection}
\begin{table}
    \centering
    \caption{Static source code metrics and static analysis warning severities used in this study including the expected direction of their values in quality increasing commits.}\label{tbl:metrics}
    \begin{tabular}{p{9cm}ll}
    \toprule
    Name and Description & Abbrev & $\updownarrow$\\
    \midrule
    Cyclomatic Complexity~\citep{mccabe}\\ The number of independent control-flow paths. & McCC & $\downarrow$\\
    \cmidrule(rl){1-3}
    Logical Lines of Code\\Number of lines in a file without comments and empty lines. & LLOC & $\downarrow$\\
    \cmidrule(rl){1-3}
    Nesting Level else-if\\Maximum of nesting level in a file. & NLE & $\downarrow$\\
    \cmidrule(rl){1-3}
    Number of parameters in a method\\The sum of all parameters of all methods in a file. & NUMPAR & $\downarrow$\\
    \cmidrule(rl){1-3}
    Clone Coverage\\Ratio of code covered by duplicates. & CC & $\downarrow$\\
    \cmidrule(rl){1-3}
    Comment lines of code\\Sum of commented lines. & CLOC & $\uparrow$\\
    \cmidrule(rl){1-3}
    Comment density\\Ratio of CLOC to LLOC. & CD & $\uparrow$\\
    \cmidrule(rl){1-3}
    API Documentation\\Number of documented public methods, +1 if class is documented. & AD & $\uparrow$\\
    \cmidrule(rl){1-3}
    Number of Ancestors\\Number of classes, interfaces, enums from which the class is inherited. & NOA & $\downarrow$\\
    \cmidrule(rl){1-3}
    Coupling between object classes\\Number of used classes (inheritance, function call, type reference). & CBO & $\downarrow$\\
    \cmidrule(rl){1-3}
    Number of Incoming Invocations\\Other methods that call the current class. & NII & $\downarrow$\\
    \cmidrule(rl){1-3}
    Minor static analysis warnings\\E.g., brace rules, naming conventions. & Minor & $\downarrow$\\
    \cmidrule(rl){1-3}
    Major static analysis warnings\\E.g., type resolution rules, unnecessary/unused code rules. & Major & $\downarrow$\\
    \cmidrule(rl){1-3}
    Critical static analysis warnings\\E.g., equals for string comparison, catching null pointer exceptions. & Critical & $\downarrow$\\
    \bottomrule
    \end{tabular}
\end{table}
The metric selection is based on the Columbus software quality model by \cite{Bakota2011}. The metrics are selected from the current version of the model also in use as QualityGate~\citep{Bakota2014}.
The current model consists of 14 static source code metrics related to size, complexity, documentation, re-usability and fault-proneness.
While the quality model provides us with a selection of metrics, we do not use it directly as it requires a baseline of projects before estimating quality of a candidate project.

Table~\ref{tbl:metrics} shows the metrics utilized in this study, a short description, and the direction which we assume they change in quality improving commits.
As most of the metrics are size and complexity metrics, we expect that their values decrease in comparison to all other commits.
The metrics we expect to increase in quality improving commits are commented lines of code, comment density, and API documentation, as added documentation should increase these metrics.
The three bottom rules consist of static analysis warnings from PMD\footnote{https://pmd.github.io/} aggregated by severity for every file.
We are of the opinion that this selection strikes a good balance of size, complexity, documentation, clone, and coupling based metrics.

As we are interested in static source code metrics in a commit granularity, we sum the metrics values for all files that are changed within a commit. In addition, we extract meta information about each change. The static source code metrics are provided by a SmartSHARK plugin using the OpenStaticAnalyzer\footnote{https://openstaticanalyzer.github.io/}.
To answer our research question, we provide the delta of the metric value changes as well as their current and previous value.

\subsection{Analysis Procedure}
\label{sec:analysis_procedure}

For our confirmatory study as part of \textbf{RQ1}, we compare the difference between two samples. To choose a valid statistical test of whether there is a difference between both samples, we first
perform the Shapiro-Wilk test~\citep{shapiro_wilk} to test for normality of each sample.
Since we found that the data is non-normal, we perform the Mann-Whitney U-test~\citep{mwu} to evaluate if the metric values of one population dominates the other.
Since we have an expectation about the direction of metric changes, we perform a one-sided Mann-Whitney U test.
The $H_0$ hypothesis is that both samples are the same, the alternative hypothesis is that one sample contains lower or higher values depending on our expectation. The expected direction of the metric value change is noted in the last column of Table~\ref{tbl:metrics}.

As our data contains a large number of metrics, we cannot assume a statistical test with $p<0.05$ is a valid rejection of a $H_0$ hypothesis.
To mitigate the problem posed by a high number of statistical tests, we perform Bonferroni correction~\citep{bonferroni}.
We choose a significance level of $\alpha=0.05$ with Bonferroni correction for 192 statistical tests.
They consist of four size metrics with two groups and three statistical tests as well as 14 source code metrics with two groups and three statistical tests (normality tests for two samples and Mann-Whitney U for difference between samples). The second part is repeated for \textbf{RQ2}.
We reject the $H_0$ hypothesis that there is no difference between samples at $p<0.00026$.

To calculate the effect size of the Mann-Whitney U test, we use Cliff's $d$~\citep{cliffsd} as a non-parametric effect size measure.
We follow a common interpretation of $d$ values~\cite{effectsizes}: $d < 0.10$ is negligible, $0.10 \leq d < 0.33$ is small, $0.33 \leq d < 0.474$ is medium and $d \ge 0.474$ is large.
We provide the effect size for every difference that is statistically significant.

We report the results visually with box plots. The box plots shows three groups: all, perfective and corrective, this allows us to show the values for each metric for each group and serves to highlight the differences.
Additionally, we report the differences between each group and its counterpart, e.g., perfective and not perfective in the tables where we report the statistical differences.

A more detailed description of the procedure for each hypothesis follows.
For \textbf{H1}, we compare the structure of quality improving changes with every other change.
We compare the size (changed lines) and diffusion (number of hunks, number of changed files) to evaluate the hypothesis.
We visualize the results with box plots and report results for statistical tests to determine if the difference in samples is statistically significant.

For \textbf{H2}, we also visualize the results via box plots. As most of the differences hover around zero, we transform the data before plotting via $sign(x)\cdot\log(abs(x + 1))$.
As we are interested in the differences between changes of metric values, we also require $x \neq 0 : \forall x \in X$ where $X$ is the complete, non-transformed data set for the visualizations.
Due to the difference in changes, we provide our data size corrected, e.g., the delta of McCC is divided by the modified lines.
Additionally, we report the percentage of data that is non-zero to indicate how often the measurements are changing in our data.
In addition to the visualization, we provide a table with differences between the samples and statistical test results.

As part of our exploratory study for answering \textbf{RQ2}, we also provide box plots of our metric values.
Instead of transformed delta values, we provide the raw averages per file in a change before the change was applied.
In addition, we provide the median values of all of our metrics before the change was applied. In this part, we apply a two-sided Mann-Whitney U test as we have no expectation of the direction the metrics change into for the categories.
To complement the visualization, we also provide density plots for both categories. They show the overlap between the perfective and corrective changes.

\subsection{Replication Kit}
All data and source code can be found in our replication kit~\citep{replication}.
In addition, we provide a small website for this publication that contains all information and where the fine-tuned model can be tested live\footnote{https://user.informatik.uni-goettingen.de/\textasciitilde{}trautsch2/emse\_2021/}.

\section{Results}\label{sec:results}
In this section, we first present the results for evaluating our hypotheses of our first research question.
After that, we describe the results of the exploratory part of our study for our second research question.

\subsection{Confirmatory Study}

We first present the results of our confirmatory study and evaluate our hypotheses.
These results answer our first research question: Does developer intent to improve internal or external quality have a positive impact on software metric values?

\clearpage
\subsubsection{Results H1}
\begin{table}
    \caption{Statistical test results for perfective and corrective commits, Mann-Whitney U test p-values ($p$-value) and effect size ($d$) with category, $n$ is negligible, $s$ is small. Statistically significant $p$-values are bolded.}\label{tbl:rq1_stats}
    \centering
    \begin{tabular}{lrr|rr}
        & \multicolumn{2}{c}{Perfective} & \multicolumn{2}{c}{Corrective}\\
        \toprule
        Metric & $p$-value & $d$ & $p$-value & $d$\\
        \midrule
\#lines added & \textbf{\textless0.0001} & 0.20 (s) & \textbf{\textless0.0001} & 0.21 (s)\\
\#lines deleted & \textbf{\textless0.0001} & 0.15 (s) & \textbf{\textless0.0001} & 0.16 (s)\\
\#files modified & 0.2081 & - & \textbf{\textless0.0001} & 0.22 (s)\\
\#hunks & \textbf{\textless0.0001} & 0.01 (n) & \textbf{\textless0.0001} & 0.22 (s)\\
        \bottomrule
    \end{tabular}
\end{table}
\begin{figure}
    \includegraphics[width=0.25\textwidth]{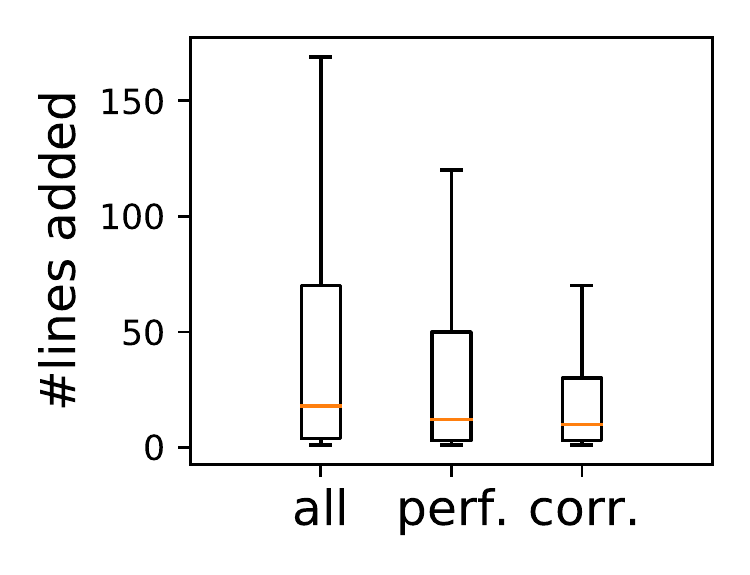}\includegraphics[width=0.25\textwidth]{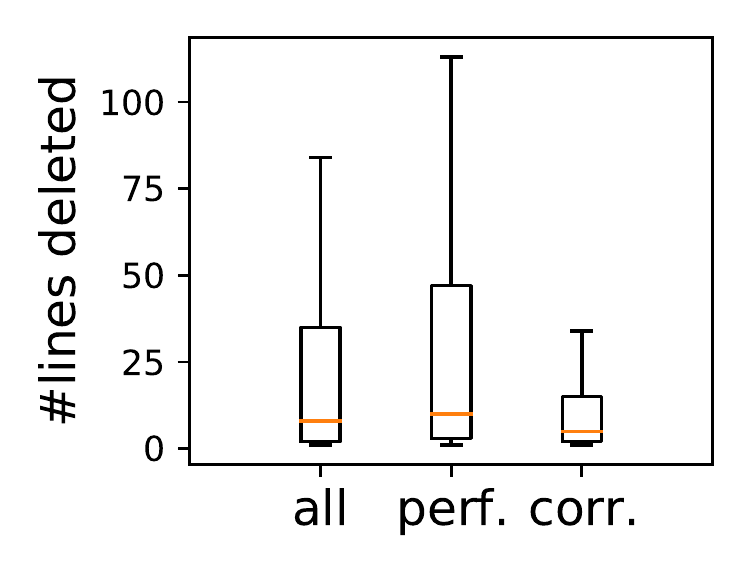}
    \includegraphics[width=0.25\textwidth]{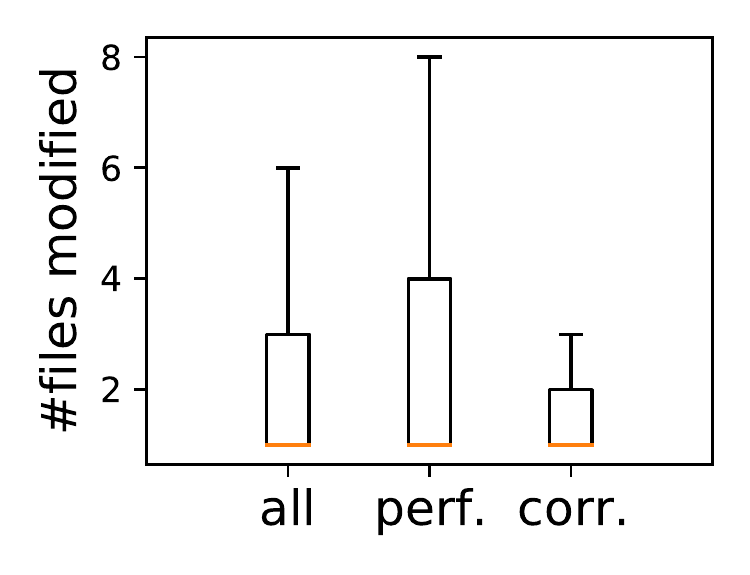}\includegraphics[width=0.25\textwidth]{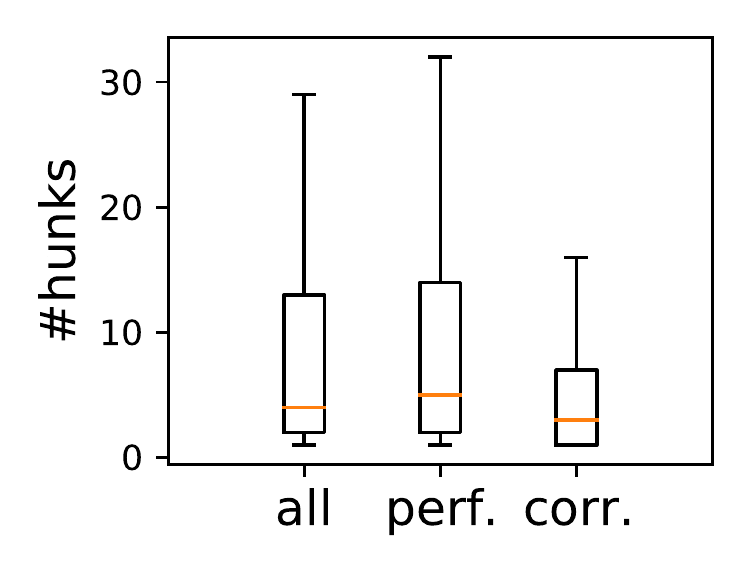}
    \caption{Commit size distribution over all projects for all, perfective and corrective commits. Fliers are omitted.}
    \label{fig:bp_dist}
\end{figure}
Figure~\ref{fig:bp_dist} shows the distribution of sizes between perfective, corrective, and all commits.
Table~\ref{tbl:rq1_stats} shows the statistical test results for the differences.
We can see that perfective commits tend to add \del{less}\rev{fewer} lines but instead remove more lines as the other commits.
When we calculate a median delta between all commits and perfective commits, we find a difference of 28 for added lines and -2 for deleted lines.
While the effect sizes are negligible to small, we can see this difference also in Figure~\ref{fig:bp_dist}.
The diffusion of the change over files is also different, however for the number of modified files the difference is not significant for perfective commits.

Corrective commits also tend to add less code, while they do not delete as much, the difference in added and deleted lines is also statistically significant.
While the effect size is small, we can see the difference in Figure~\ref{fig:bp_dist}.
For corrective commits, we can also see a difference in the number of files changed and the number of hunks modified.
This diffusion of the change via the number of files and hunks is also statistically significant although, again, with a small effect size.

We can conclude, that perfective commits tend to remove more lines, and are generally adding \del{less}\rev{fewer} lines to the repository.
Corrective commits delete \del{less}\rev{fewer} lines and add \del{less}\rev{fewer} lines than other commits.
Corrective commits are also distributed over less hunks and less files than other commits.

\vspace{1em}
\begin{center}
    \setlength{\fboxrule}{0pt}
        \vrule\vrule\vrule\vrule\vrule\vrule\fcolorbox{white}{gray!15}{
            \hspace{.2em}\parbox{.91\linewidth} {
            \vspace{.5em}
        We \textbf{accept H1} that intended quality improvements are smaller than other changes.\\
        Perfective and corrective commits tend to add\del{less}\rev{fewer} lines, perfective commits remove more lines.
        The effect size is negligible to small in all cases.
            \vspace{.5em}
        }\hspace{.5em}
    }
\end{center}

\vspace{.5em}

\subsubsection{Results H2}
We first note that no metric value changes for each instance of our data.
This can be seen in Table~\ref{tbl:commits}, which shows the percentages for each metric value for perfective, corrective, and all changes.
\begin{table}
    \caption{Percentage of commits where the metric value is not zero on all commits (\%nz), perfective commits (\%nz p) and corrective commits (\%nz c).}
    \label{tbl:commits}
    \centering
    \begin{tabular}{lrrr}
	    Metric & \%NZ & \%NZ P & \%NZ C\\
        \toprule
McCC & 51.03 & 31.01 & 57.70\\
LLOC & 74.69 & 60.93 & 77.99\\
NLE & 36.76 & 23.92 & 34.28\\
NUMPAR & 35.93 & 24.44 & 24.98\\
CC & 49.41 & 37.81 & 55.14\\
CLOC & 51.56 & 46.52 & 42.51\\
CD & 76.07 & 66.48 & 77.35\\
AD & 27.19 & 20.63 & 15.82\\
NOA & 10.51 & 6.96 & 3.62\\
CBO & 30.89 & 22.52 & 22.22\\
NII & 27.08 & 17.78 & 21.09\\
Minor & 36.15 & 27.02 & 29.77\\
Major & 19.87 & 13.23 & 14.77\\
Critical & 7.23 & 4.20 & 4.95\\
        \bottomrule
	\end{tabular}
\end{table}
We can see some differences between changes, e.g., critical PMD warnings only change in about 7\% of commits while LLOC changes in about 75\%.
Some differences are also between categories, e.g., McCC changes in 31\% of perfective changes and in 57\% of corrective changes.

\begin{figure}
    \includegraphics[width=0.24\textwidth]{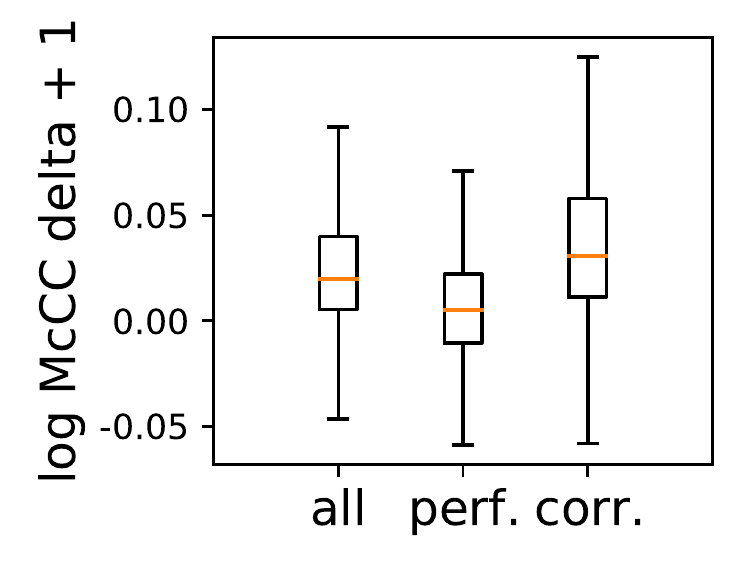}\includegraphics[width=0.24\textwidth]{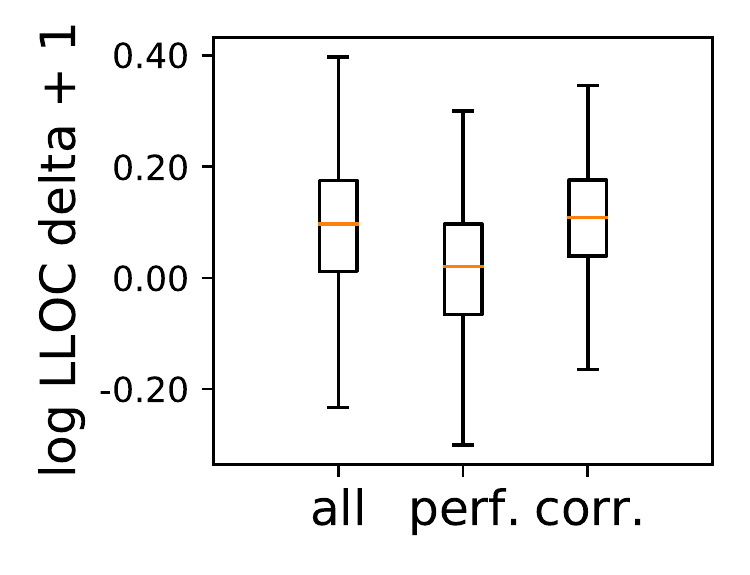}
    \includegraphics[width=0.24\textwidth]{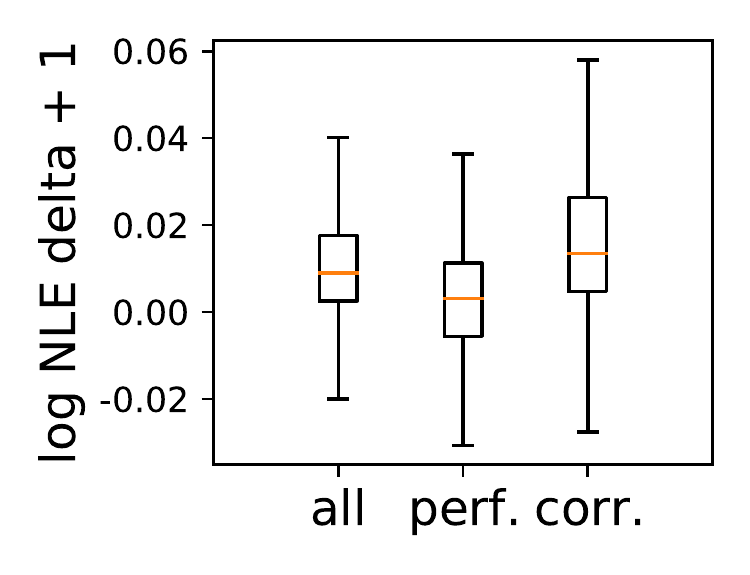}\includegraphics[width=0.24\textwidth]{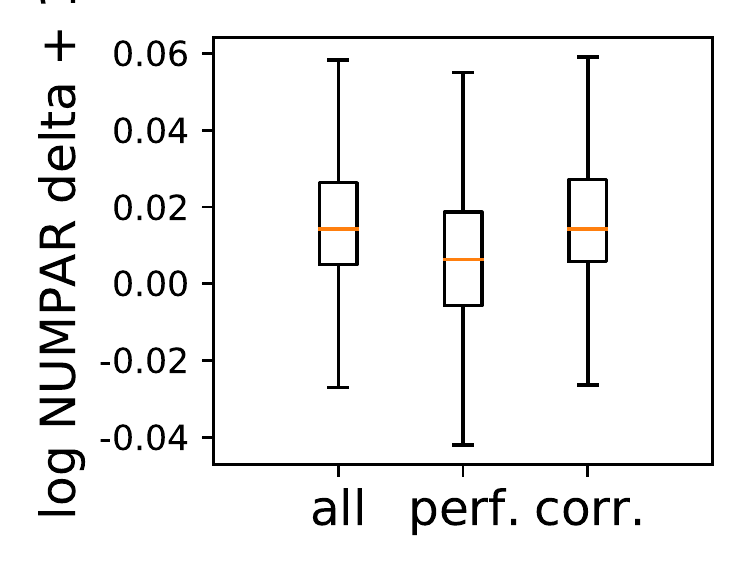}
    \includegraphics[width=0.24\textwidth]{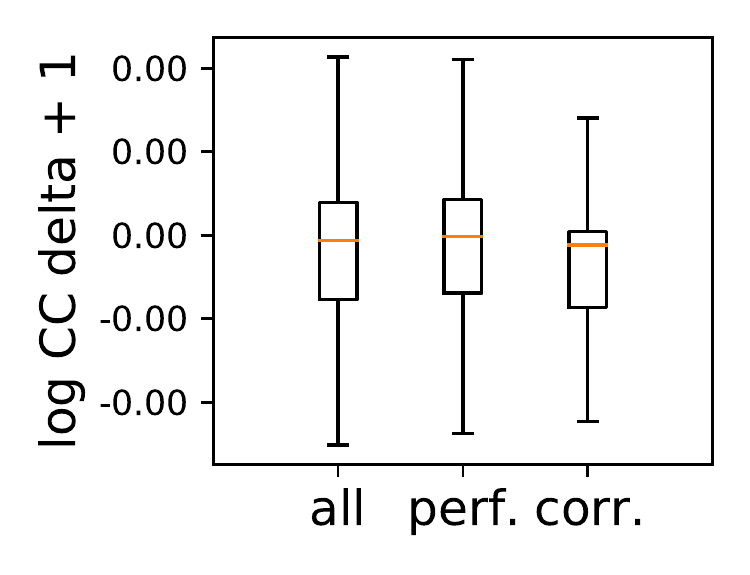}\includegraphics[width=0.24\textwidth]{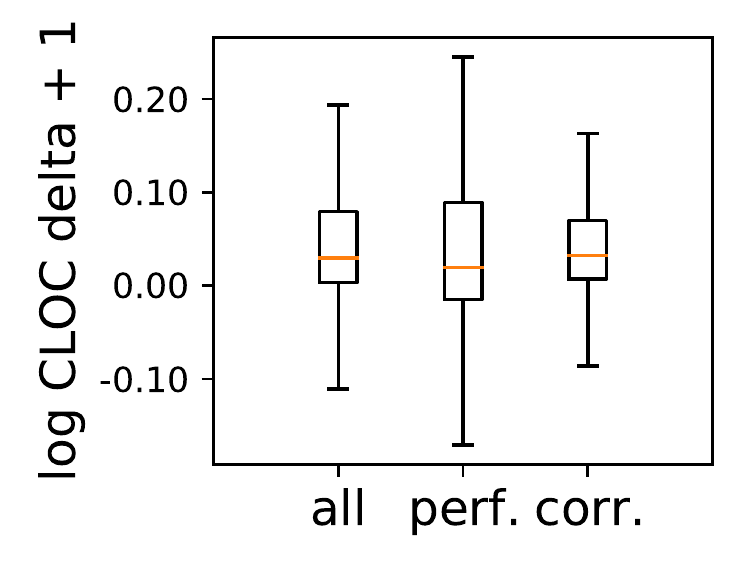}
    \includegraphics[width=0.24\textwidth]{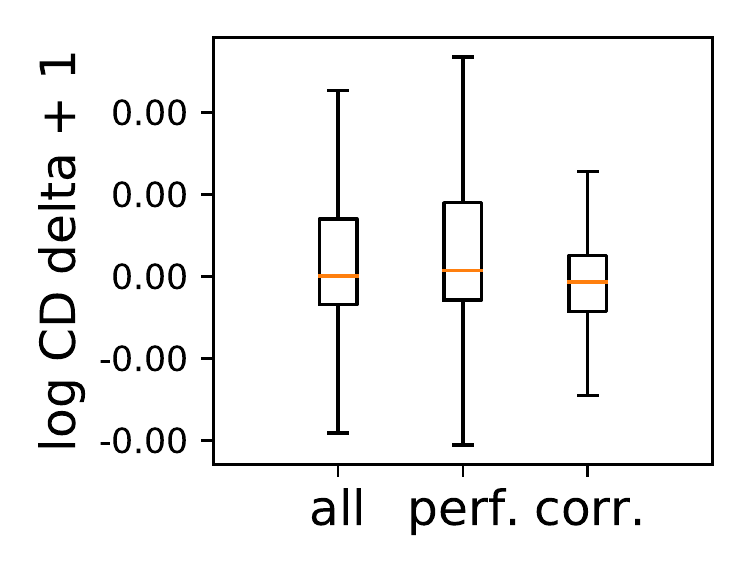}\includegraphics[width=0.24\textwidth]{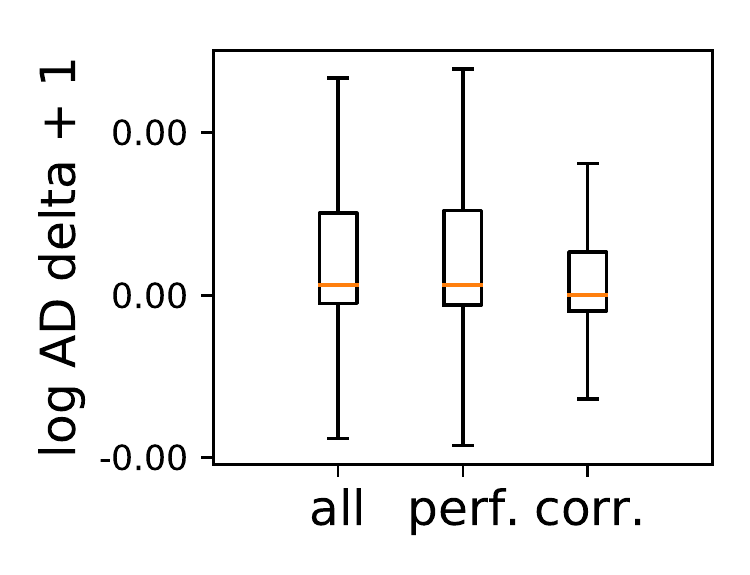}
    \includegraphics[width=0.24\textwidth]{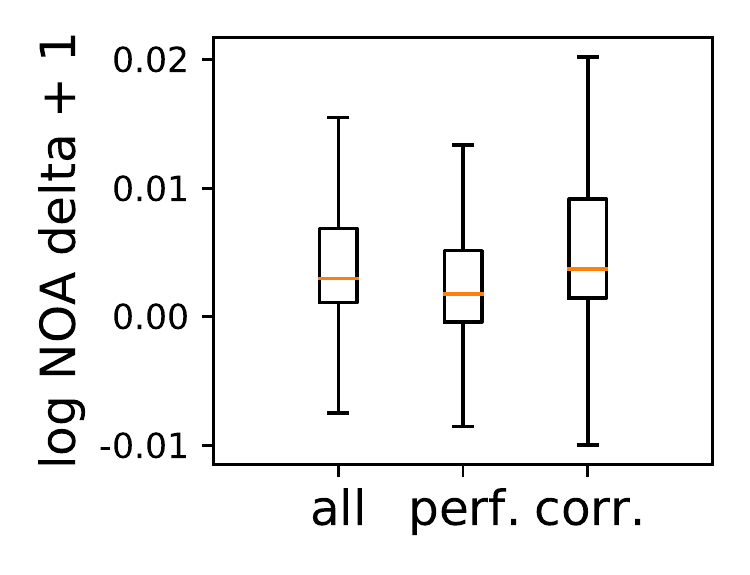}\includegraphics[width=0.24\textwidth]{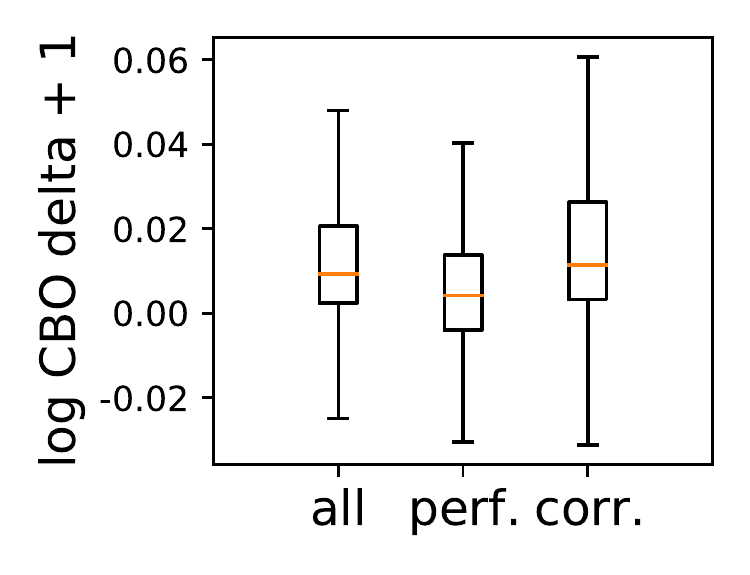}
    \includegraphics[width=0.24\textwidth]{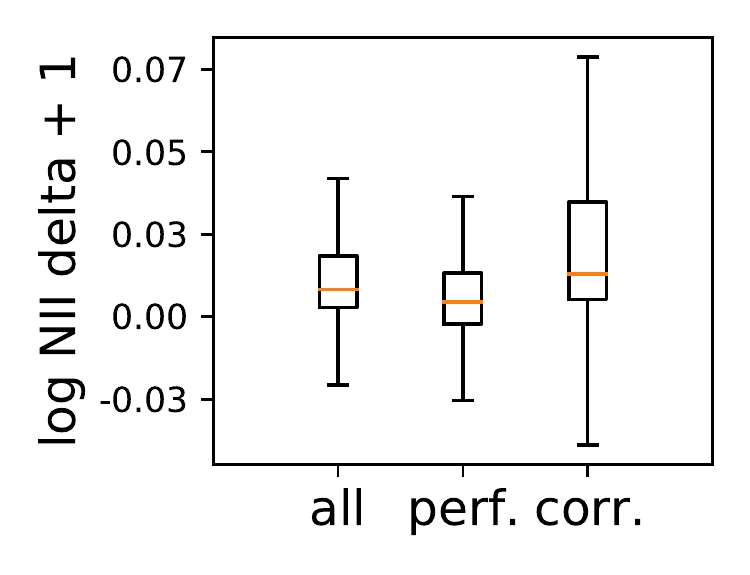}\includegraphics[width=0.24\textwidth]{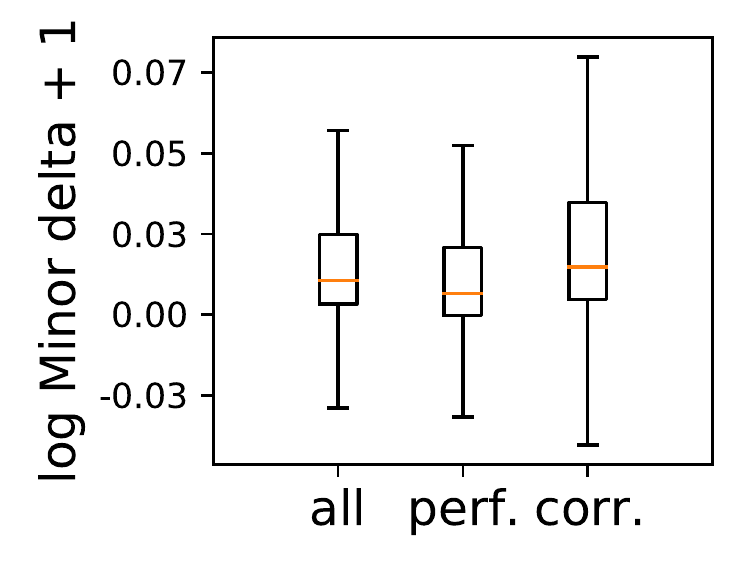}
    \includegraphics[width=0.24\textwidth]{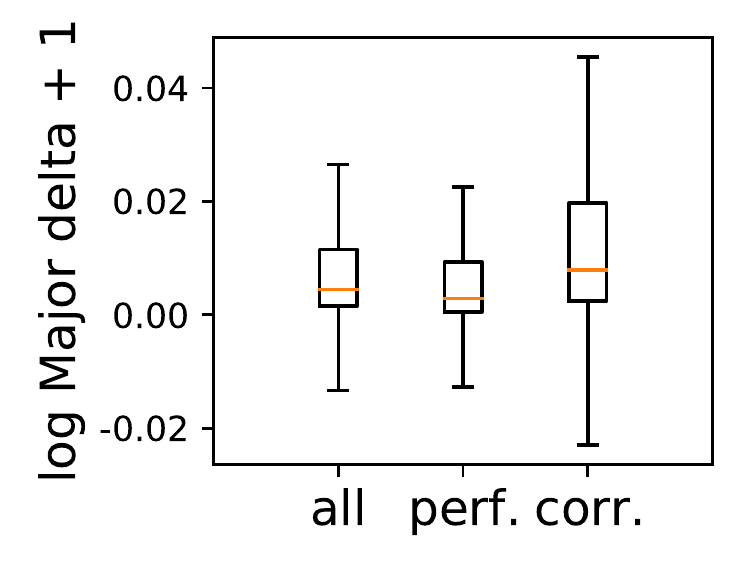}\includegraphics[width=0.24\textwidth]{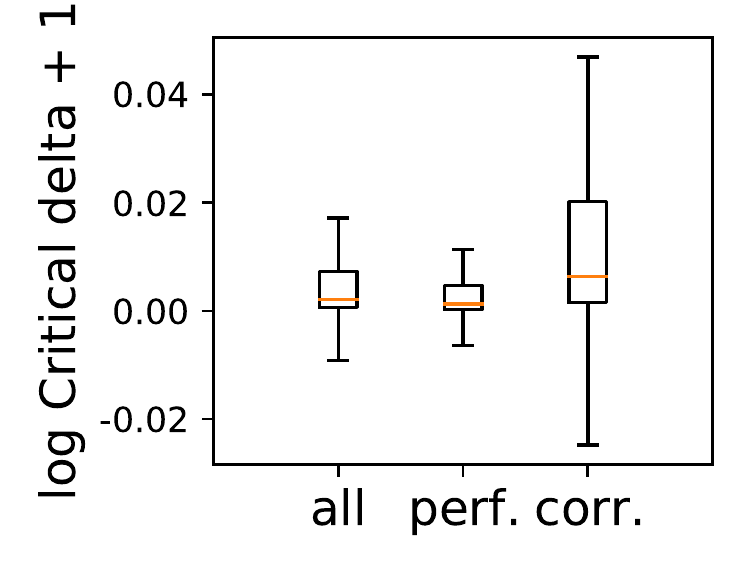}
    \caption{Static source code metric value changes in all, perfective and corrective commits divided by changed lines. Fliers are omitted.}
    \label{fig:bp_static_density}
\end{figure}

To evaluate \textbf{H2}, we present the differences in all changes visually as box plots in Figure~\ref{fig:bp_static_density}.

\begin{table}
    \caption{Statistical test results for perfective and corrective commits, Mann-Whitney U test p-values ($p$-value) and effect size ($d$) with category, $n$ is negligible, $s$ is small, $m$ is medium. Statistically significant $p$-values are bolded. All values are normalized for changed lines.}\label{tbl:diff_lloc}
    \centering
    \begin{tabular}{lrr|rr}
        & \multicolumn{2}{c}{Perfective} & \multicolumn{2}{c}{Corrective}\\
        \toprule
        Metric & $p$-val & $d$ & $p$-val & $d$\\
        \midrule
McCC & \textbf{\textless0.0001} & 0.39 (m) & 1.0000 & -\\
LLOC & \textbf{\textless0.0001} & 0.45 (m) & 1.0000 & -\\
NLE & \textbf{\textless0.0001} & 0.27 (s) & 1.0000 & -\\
NUMPAR & \textbf{\textless0.0001} & 0.25 (s) & \textbf{\textless0.0001} & 0.09 (n)\\
CC & 1.0000 & - & \textbf{\textless0.0001} & 0.12 (s)\\
CLOC & \textbf{\textless0.0001} & 0.16 (s) & \textbf{\textless0.0001} & 0.05 (n)\\
CD & 1.0000 & - & \textbf{\textless0.0001} & 0.16 (s)\\
AD & \textbf{\textless0.0001} & 0.02 (n) & \textbf{\textless0.0001} & 0.08 (n)\\
NOA & \textbf{\textless0.0001} & 0.08 (n) & \textbf{\textless0.0001} & 0.07 (n)\\
CBO & \textbf{\textless0.0001} & 0.19 (s) & \textbf{\textless0.0001} & 0.06 (n)\\
NII & \textbf{\textless0.0001} & 0.19 (s) & \textbf{\textless0.0001} & 0.02 (n)\\
Minor & \textbf{\textless0.0001} & 0.19 (s) & \textbf{\textless0.0001} & 0.05 (n)\\
Major & \textbf{\textless0.0001} & 0.12 (s) & \textbf{\textless0.0001} & 0.05 (n)\\
Critical & \textbf{\textless0.0001} & 0.05 (n) & \textbf{\textless0.0001} & 0.03 (n)\\
        \bottomrule
    \end{tabular}
\end{table}
In addition, we provide Table~\ref{tbl:diff_lloc} which shows the Mann-Whitney U test~\citep{mwu} p-values, and effect sizes for differences between the types of commits.
We can see that most metric values are different depending on whether they are measured in perfective, corrective, or all other commits.
In the following, we discuss the differences for each measured metric value. A description for each metric and the expected direction of metric value change is shown in Table~\ref{tbl:metrics}.

\textbf{McCC}: the cyclomatic complexity of perfective changes is smaller than for other changes. Even when we do not account for the size of the change.
This is expected as some perfective commits mention simplification of code. For perfective commits the effect size is medium.
Corrective commits however have higher McCC than other commits. This can be seen in Figure~\ref{fig:bp_static_density}. The median of corrective commits is higher than for the other commits.
Our assumption about McCC being lower in all quality improving commits is not met in this case.
While it makes sense that corrective commits add complexity, the comparison here is one of stochastic dominance between all other commits
and only corrective commits, not if corrective commits remove or add McCC.
Thus, this means that changes in corrective commits are more complex than those of other changes.

\textbf{LLOC}: the difference of LLOC is the most pronounced in our data. We find that even when we do not correct for size of the change the difference between perfective and other changes in LLOC is the most pronounced.
While manually classifying the commits, we found that often code is removed because it was marked as deprecated before or it was no longer needed due to other reasons. The effect size for perfective commits is medium.
For corrective commits, we can see the same result as for McCC. While we assumed that bug fixes usually add code, we did not expect them to dominate all other commits including feature additions.

\textbf{NLE}: the nesting level if-else is smaller in perfective commits. We expect this is due to simplification and removal of complex code. When we look at the box plot in Figure~\ref{fig:bp_static_density} it shows a noticeable difference.
This means simplification is a high priority when improving code quality in perfective commits.
For corrective commits, we can see the same effect as previously seen for McCC and LLOC. The NLE is not lower but higher for corrective commits. This is more evidence for the fact that bug fixes add more complex code. There may be a timing factor involved, e.g., if bug fixes are quick fixes, they would add more complex code without a more complex refactoring which would decrease the complexity again.

\textbf{NUMPAR}: the number of parameters in a method is also different for perfective commits. This may be a hint of the type of perfective maintenance performed the most in perfective commits.
The manual classification showed a lot of commit messages that claimed a simplification of the changed code. This metric would also be impacted by a simplification or refactoring operations.
Corrective commits also show less additions in this metric, while it only has a negligible effect size it is still statistically significant.
Fixing bugs seems to include some code reduction or at least less addition of parameters for methods.

\textbf{CC}: the clone coverage is not different for perfective commits.
We would have expected that it is decreasing in perfective commits. However, it seems that clone removal is not a big part of perfective maintenance in our study subjects, which contradicts our expectation.
Corrective commits contain a lower clone coverage\rev{,} however.
This could either be because corrective commits introduce fewer new clones than other commits or because they remove more. A possible reason for clone removal may be the correction of copy and paste related bugs.

\textbf{CLOC}: the comment lines of code show a difference for perfective commits and corrective commits.
While we expected the CLOC to increase in both types of quality improving commits the effect size is higher in perfective commits.
It seems that bug fixing operations do not add enough comment lines to show a larger difference here for corrective commits.

\textbf{CD}: the comment density of perfective commits is not statistically significantly different from other commits.
We would have expected a difference here because perfective maintenance should include additional comments on new or previously uncommented code.
We can see a difference for corrective commits here. This shows that the density of comments is also improving in bug fixing operations probably due to clarifications for parts of the code that were fixed.

\textbf{AD}: the API documentation metric does change in perfective and corrective commits compared to other commits.
A reason could be that perfective commits do add API documentation to make the difference significant.
Corrective changes that introduce code in our study subjects seem to almost always include API documentation, therefore we can see a difference here.
However, the effect size is negligible in both cases.

\textbf{NOA}: the number of ancestors is lower in perfective commits as expected. This metric would be affected in simplification and clean up maintenance operations.
For corrective commits we can also see a lower value, this hints at some clean up operations happening during bug fixing.

\textbf{CBO}: the coupling between objects is lower after perfective commits. This is expected due to class removal and subsequent decoupling of classes.
For corrective commits we can also see a difference. While the effect size is negligible, there is some code clean up happening during bug fixes, e.g., NOA and CC are also lower in corrective than in other commits.

\textbf{NII}: the number of incoming invocations is lower in both perfective and corrective commits.
However, the effect size is small in perfective and negligible in corrective commits.
It seems reasonable to see a difference in this metric, because in the case of perfective commits,
we have lots of source code removal. However, there are also maintenance activities which are decoupling classes which would also impact this metric.
Corrective maintenance seems to involve only limited decoupling operations, also seen in CBO.

\textbf{Minor}: The PMD warnings of minor severity are different in both types of changes.
However, we can see that the effect size is larger for perfective changes which makes sense as those warnings can be part of perfective maintenance.

\textbf{Major}: The PMD warnings of major severity are also different in both types of changes.
We can see the difference in effect size again and we expect the reason is the same as for Minor.

\textbf{Critical}: The PMD warnings of critical severity are different for both types of changes.
Here, the effect size is negligible for both types.
However, as they are only changed in about 7\% of our commits, they are not changing often regardless of commit type.

\vspace{.8em}

\noindent
\begin{center}
    \setlength{\fboxrule}{0pt}
        \vrule\vrule\vrule\vrule\vrule\vrule\fcolorbox{white}{gray!15}{
            \hspace{.2em}\parbox{.91\linewidth} {
                \vspace{.5em}
There are significant differences between perfective and corrective changes.
We \textbf{reject H2} that intended quality improvements have a positive impact on quality metric values.\vspace{.5em}

                \vspace{.5em}
        }\hspace{.5em}
    }
\end{center}

\subsection{Summary RQ1}

In summary, we have the following results for RQ1.
\begin{center}
    \setlength{\fboxrule}{0pt}
        \vrule\vrule\vrule\vrule\vrule\vrule\fcolorbox{white}{gray!15}{
            \hspace{.2em}\parbox{.91\linewidth} {
                \vspace{.5em}
                \textbf{RQ1 Summary}\\
                While intended quality improvements by developers yield measurable differences in almost all metrics we find that not all metric values are changing in the expected direction.\\

                \textbf{Perfective changes}\\
Perfective commits have a positive effect on metric values that measure code complexity through the size, conditional statements, number of parameters, and coupling.
For two metrics we do not find the expected difference to other commits.
Code clones and comment density metric values are not statistically significantly different in perfective commits.\\

\textbf{Corrective changes}\\
Only for two metrics, we observe a non\rev{-}negligible and statistically significant change that we predicted.
For LLOC, McCC and NLE, we observe the opposite of the expectation, which indicates that bug fixes add complex code.
                \vspace{.5em}
        }\hspace{.5em}
    }
\end{center}

\subsection{Exploratory Study}
To answer our \textbf{RQ2:} What kind of files are the target of internal or external quality improvements?
We conduct an exploratory study. We present the results which files are changed in which change category with respect to their metric values.
The extracted metrics are considered on a per-change basis, i.e., we divide the metrics by the number of changed files to get an average metric value per file.

\begin{figure}
    \includegraphics[width=0.24\textwidth]{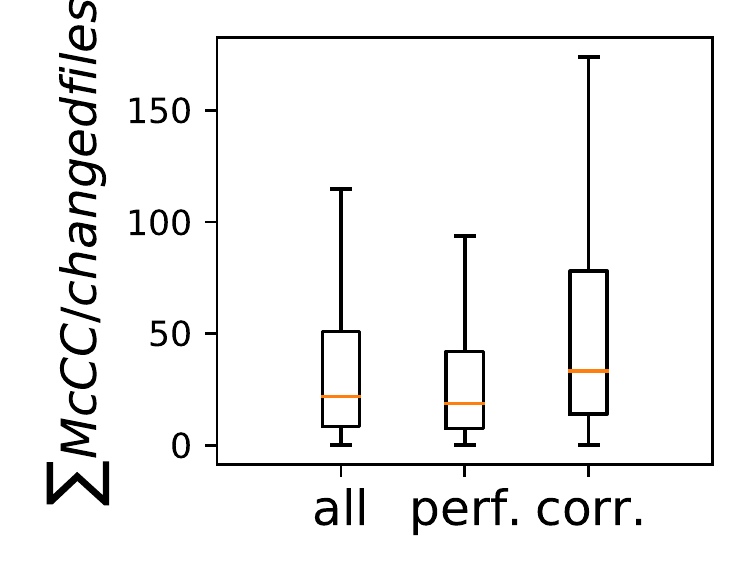}\includegraphics[width=0.24\textwidth]{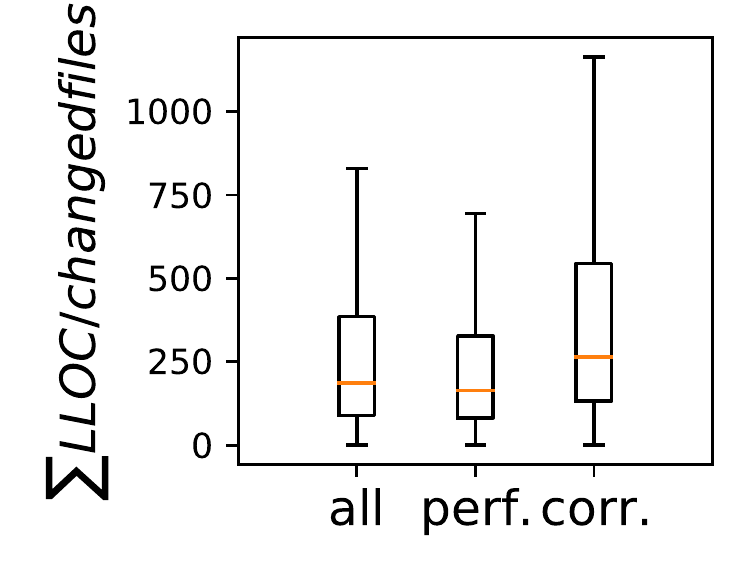}
    \includegraphics[width=0.24\textwidth]{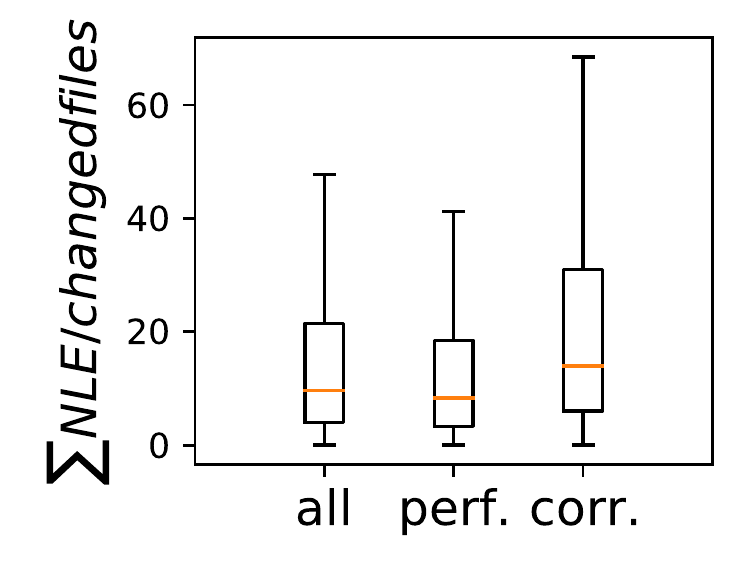}\includegraphics[width=0.24\textwidth]{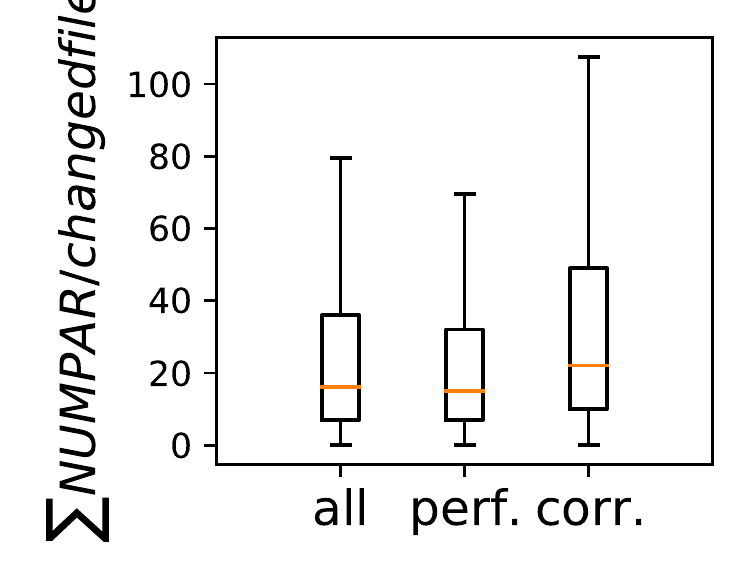}
    \includegraphics[width=0.24\textwidth]{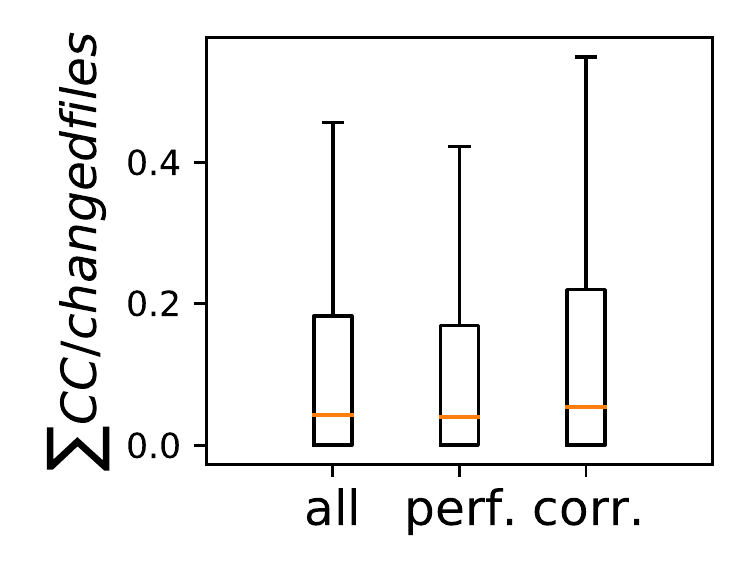}\includegraphics[width=0.24\textwidth]{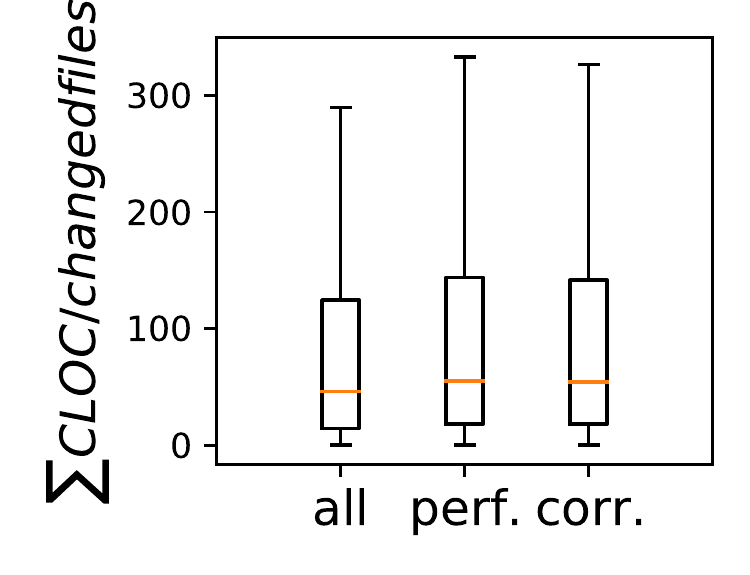}
    \includegraphics[width=0.24\textwidth]{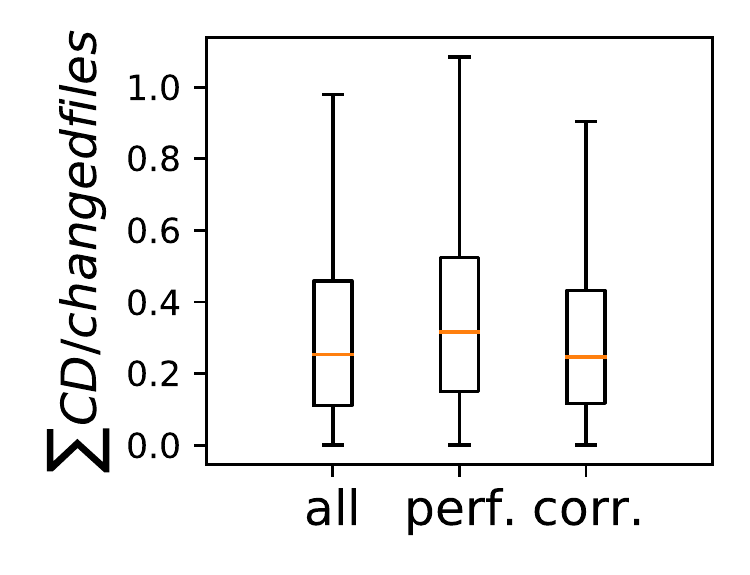}\includegraphics[width=0.24\textwidth]{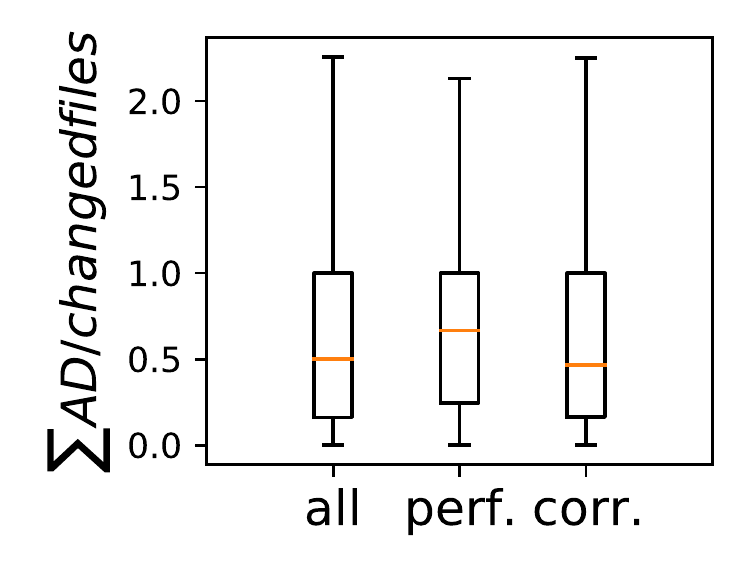}
    \includegraphics[width=0.24\textwidth]{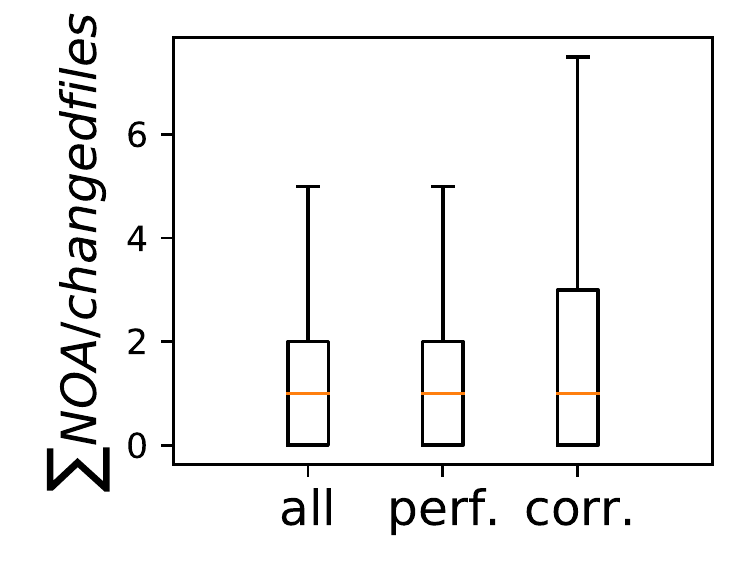}\includegraphics[width=0.24\textwidth]{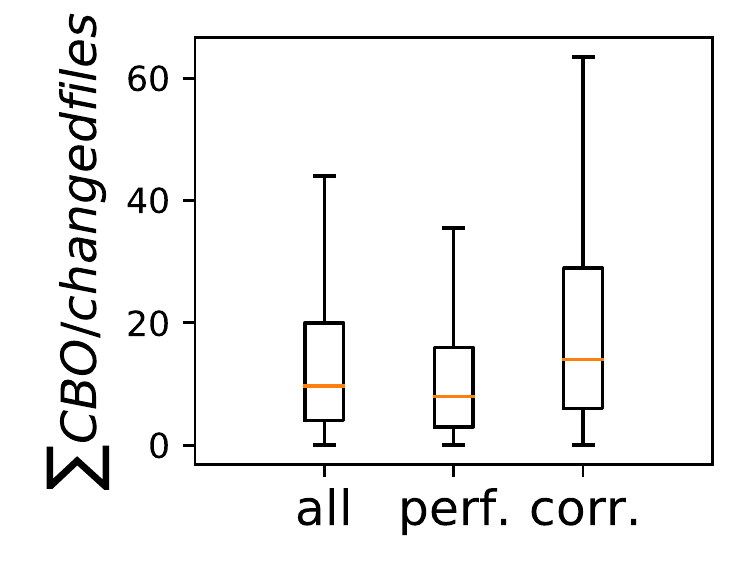}
    \includegraphics[width=0.24\textwidth]{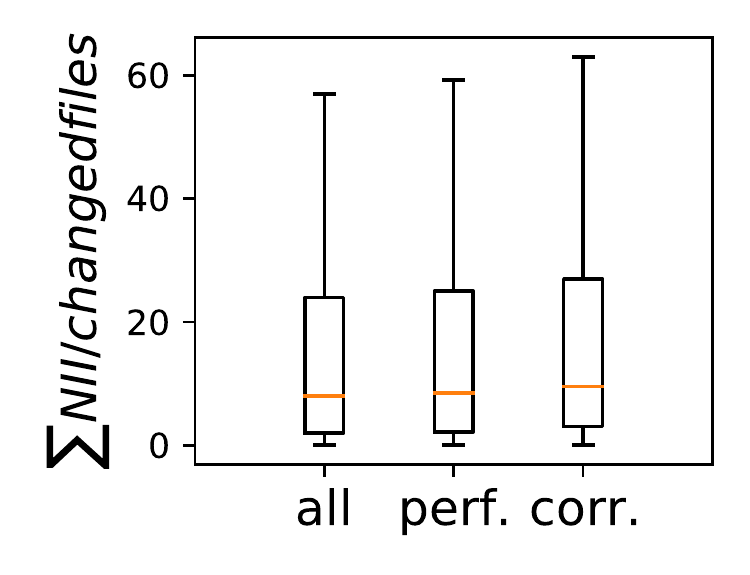}\includegraphics[width=0.24\textwidth]{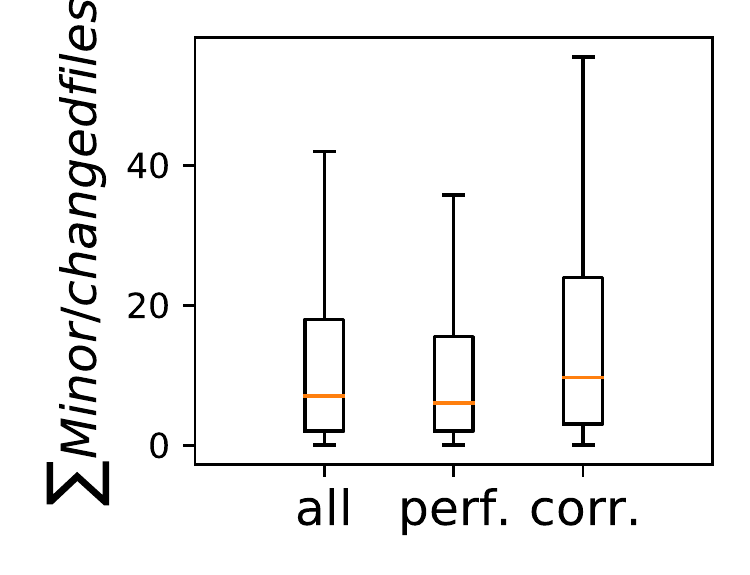}
    \includegraphics[width=0.24\textwidth]{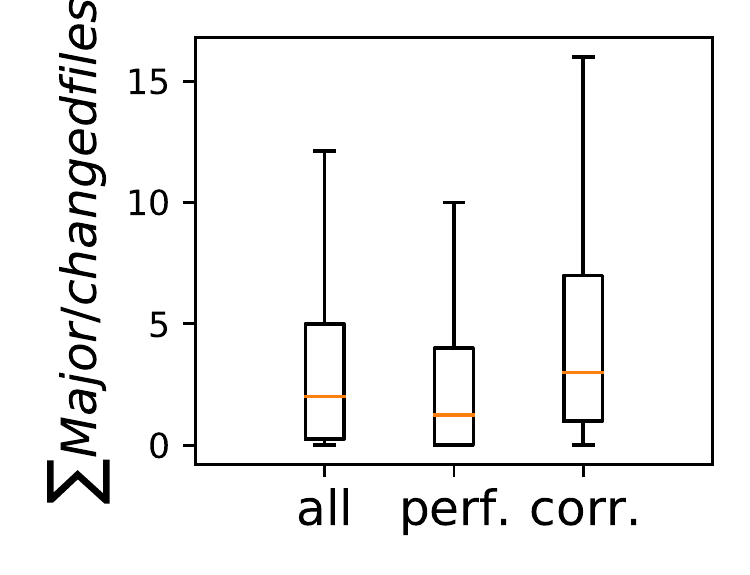}\includegraphics[width=0.24\textwidth]{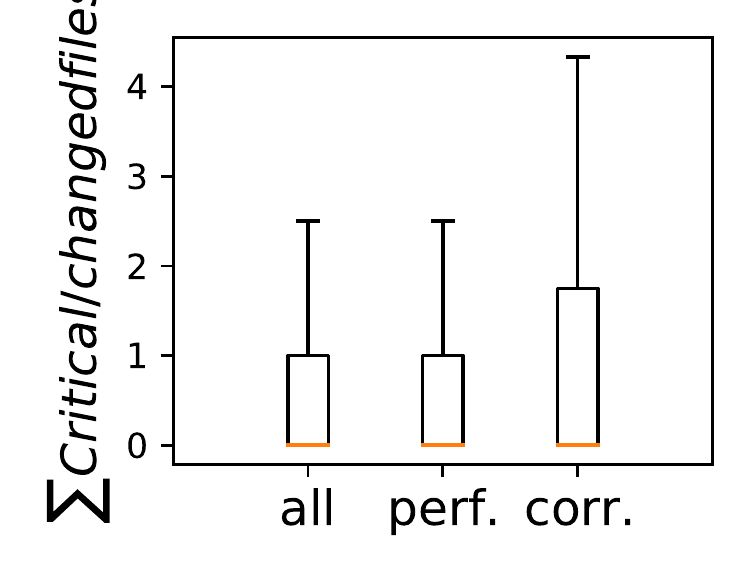}
    \caption{Static source code metrics before the change is applied. Fliers are omitted.}
    \label{fig:bp_static_parent}
\end{figure}
Figure~\ref{fig:bp_static_parent} shows box plots for the metric values of files before the change is applied. We can see that, perfective changes are not necessarily applied to complex files.
If we compare the median values in Table~\ref{tbl:median} we can see that perfective changes are applied to smaller, simpler files than the average or corrective change. McCC, LLOC, NLE, NUMPAR and CBO are lower for the files which receive perfective changes, while CLOC, CD, AD are higher. This means that less complex and well documented files are often the target of perfective changes.
If we look at corrective changes we see that they are more complex and usually larger files.
McCC, LLOC, NLE, NUMPAR, CBO, NII as well as Minor, Major and Critical are higher than other, or perfective changes.
As we consider the metric values before the change is applied they can be considered pre-bugfix. However, when we consider our results for \textbf{RQ1} the corrective changes usually increase the complexity even further.

Table~\ref{tbl:stats_parent} show the results of our statistical tests.
Analogous to \textbf{RQ1} we compare the difference between perfective and non-perfective as well as corrective and non-corrective.
While most metric differences are statistically significant, we observe only some small effect sizes for the comment related metrics while the rest is negligible.

\begin{table}
    \caption{Median metric values per file before the change is applied.}\label{tbl:median}
    \centering
    \begin{tabular}{lrrr}
        \toprule
        Metric & All & Perfective & Corrective\\
        \midrule
McCC & 21.78 & 18.78 & 33.23\\
LLOC & 186.98 & 163.75 & 264.18\\
NLE & 9.60 & 8.33 & 14.00\\
NUMPAR & 16.06 & 15.00 & 22.00\\
CC & 0.04 & 0.04 & 0.05\\
CLOC & 46.25 & 55.00 & 54.00\\
CD & 0.25 & 0.32 & 0.25\\
AD & 0.50 & 0.67 & 0.46\\
NOA & 1.00 & 1.00 & 1.00\\
CBO & 9.67 & 8.00 & 14.00\\
NII & 8.00 & 8.50 & 9.50\\
Minor & 7.00 & 6.00 & 9.67\\
Major & 2.00 & 1.25 & 3.00\\
Critical & 0.00 & 0.00 & 0.00\\
\bottomrule
    \end{tabular}
\end{table}

\begin{table}
    \caption{Statistical test results for perfective and corrective commits regarding their average metrics before the change, Mann-Whitney U test p-values ($p$-value) and effect size ($d$) with category, $n$ is negligible, $s$ is small, $m$ is medium. Statistically significant $p$-values are bolded.}\label{tbl:stats_parent}
    \centering
    \begin{tabular}{lrr|rr}
        & \multicolumn{2}{c}{Perfective} & \multicolumn{2}{c}{Corrective}\\
        \toprule
        Metric & $p$-val & $d$ & $p$-val & $d$\\
        \midrule
McCC & \textbf{\textless0.0001} & 0.05 (n) & \textbf{\textless0.0001} & 0.08 (n)\\
LLOC & \textbf{\textless0.0001} & 0.05 (n) & \textbf{\textless0.0001} & 0.05 (n)\\
NLE & \textbf{\textless0.0001} & 0.04 (n) & \textbf{\textless0.0001} & 0.07 (n)\\
NUMPAR & 0.6367 & - & 0.0218 & -\\
CC & \textbf{\textless0.0001} & 0.01 (n) & 0.0011 & -\\
CLOC & \textbf{\textless0.0001} & 0.12 (s) & \textbf{\textless0.0001} & 0.06 (n)\\
CD & \textbf{\textless0.0001} & 0.15 (s) & \textbf{\textless0.0001} & 0.15 (s)\\
AD & \textbf{\textless0.0001} & 0.17 (s) & \textbf{\textless0.0001} & 0.15 (s)\\
NOA & 0.5109 & - & \textbf{\textless0.0001} & 0.02 (n)\\
CBO & \textbf{\textless0.0001} & 0.09 (n) & \textbf{\textless0.0001} & 0.07 (n)\\
NII & \textbf{\textless0.0001} & 0.05 (n) & \textbf{\textless0.0001} & 0.04 (n)\\
Minor & \textbf{\textless0.0001} & 0.04 (n) & \textbf{\textless0.0001} & 0.02 (n)\\
Major & \textbf{\textless0.0001} & 0.09 (n) & \textbf{\textless0.0001} & 0.04 (n)\\
Critical & \textbf{\textless0.0001} & 0.05 (n) & \textbf{\textless0.0001} & 0.03 (n)\\
    \bottomrule
    \end{tabular}
\end{table}

\begin{figure}
    \includegraphics[width=0.24\textwidth]{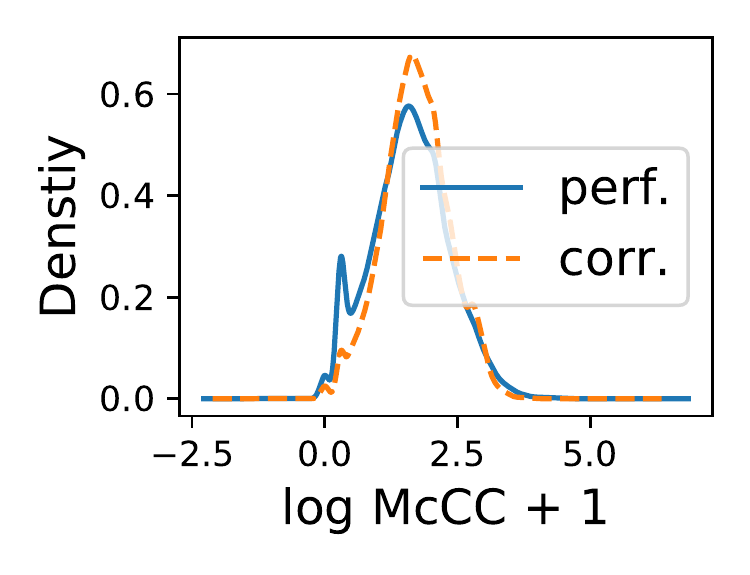}\includegraphics[width=0.24\textwidth]{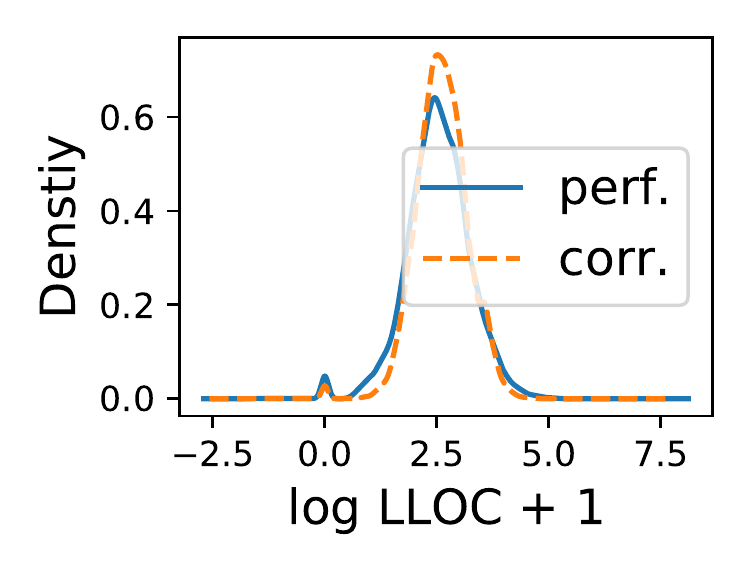}
    \includegraphics[width=0.24\textwidth]{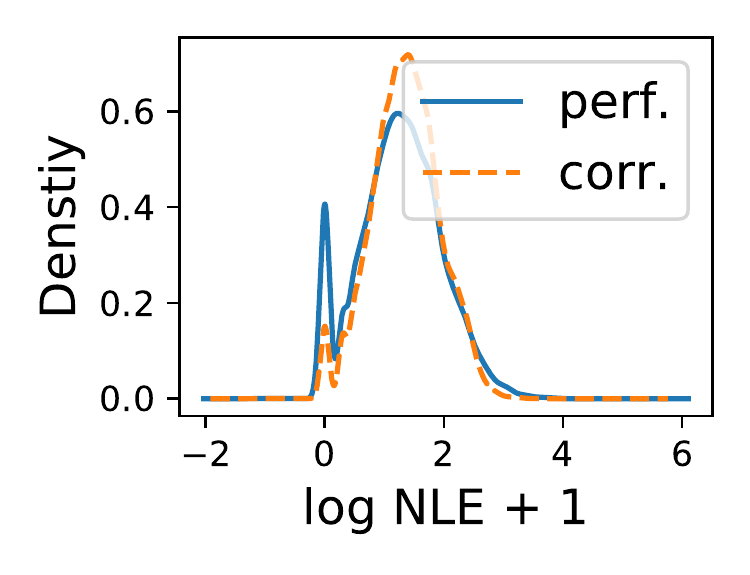}\includegraphics[width=0.24\textwidth]{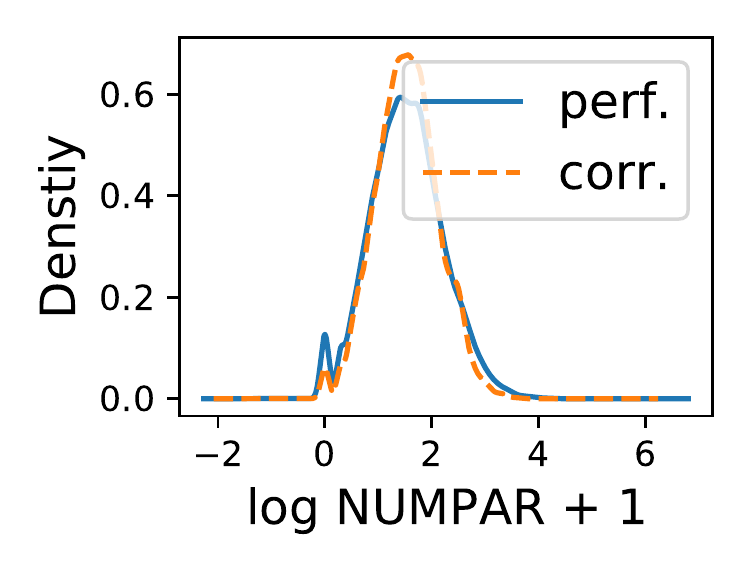}
    \includegraphics[width=0.24\textwidth]{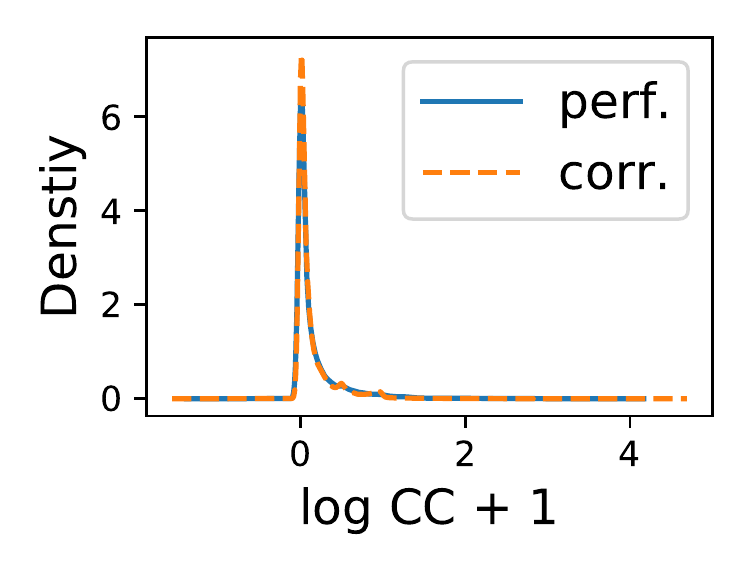}\includegraphics[width=0.24\textwidth]{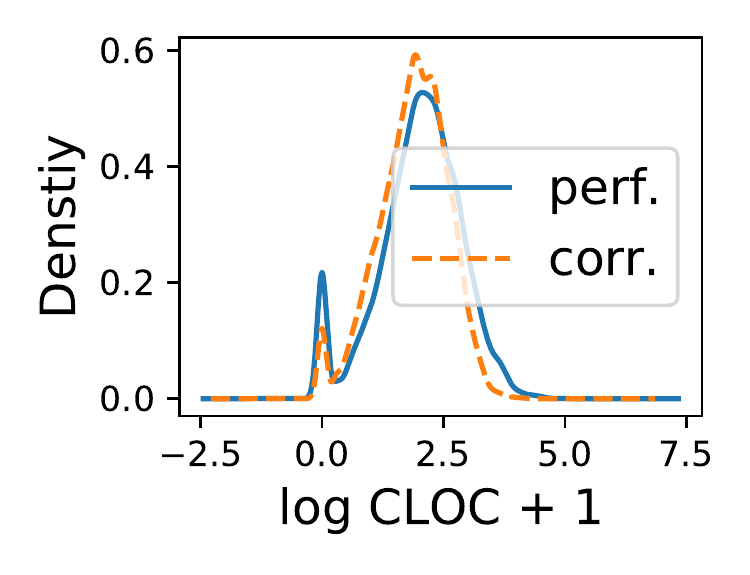}
    \includegraphics[width=0.24\textwidth]{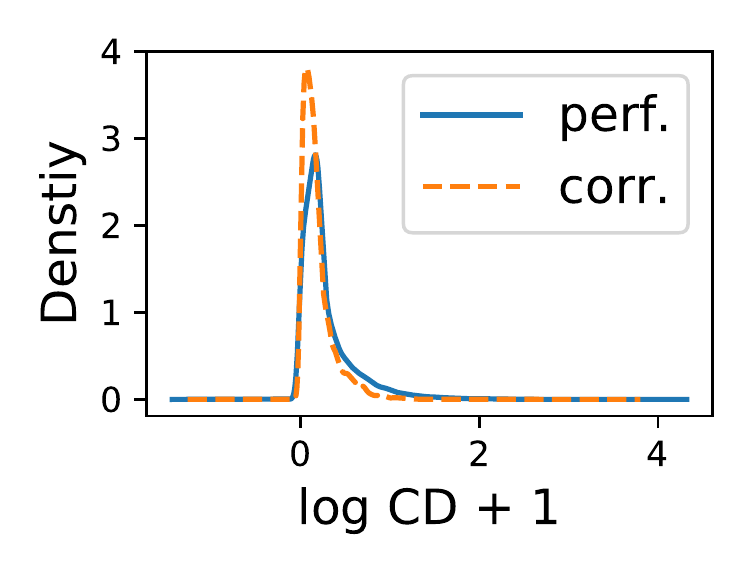}\includegraphics[width=0.24\textwidth]{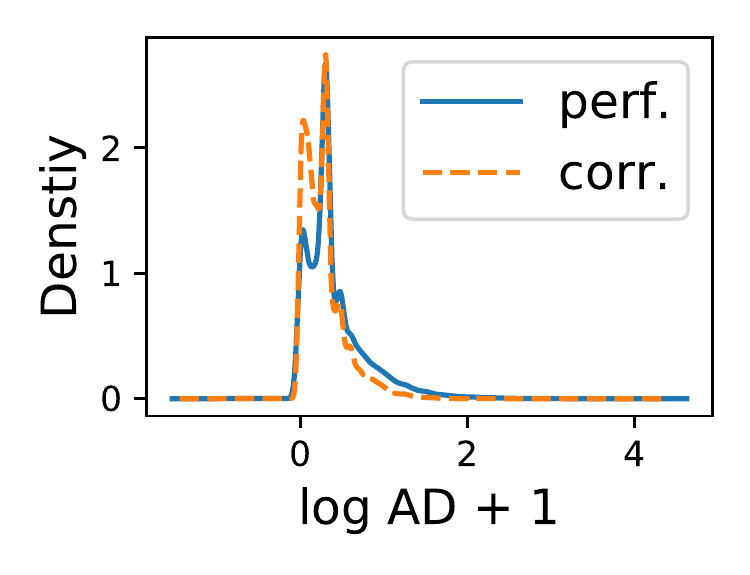}
    \includegraphics[width=0.24\textwidth]{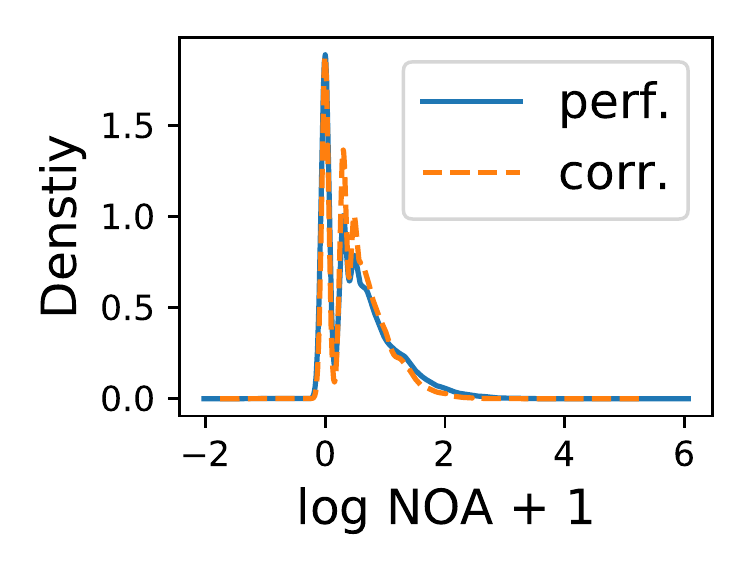}\includegraphics[width=0.24\textwidth]{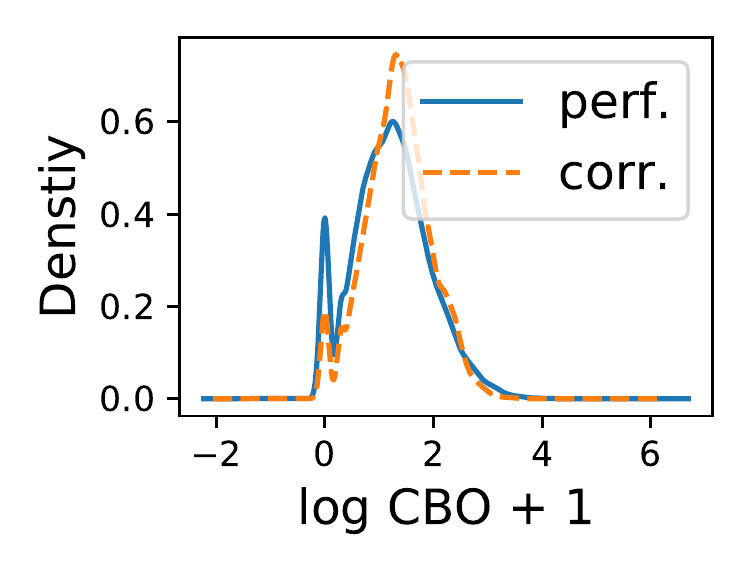}
    \includegraphics[width=0.24\textwidth]{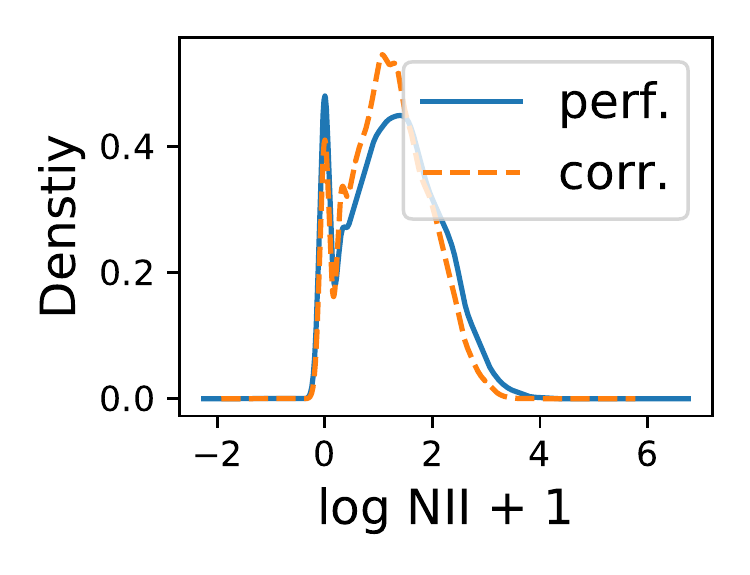}\includegraphics[width=0.24\textwidth]{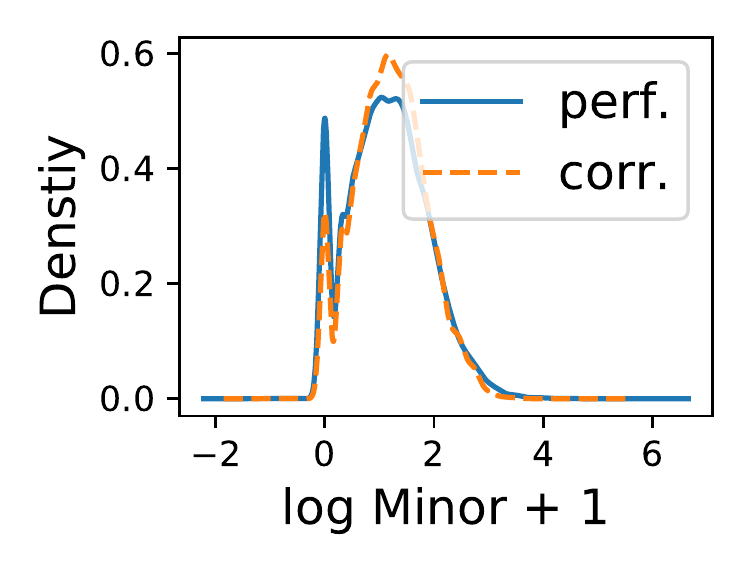}
    \includegraphics[width=0.24\textwidth]{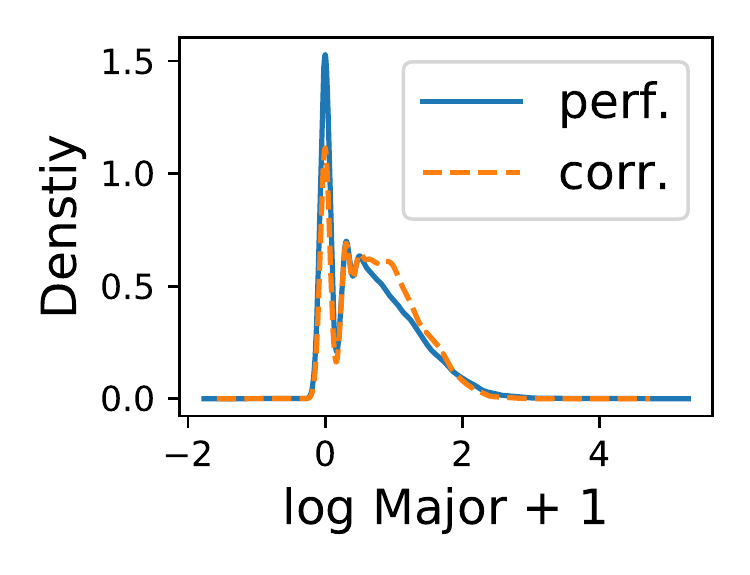}\includegraphics[width=0.24\textwidth]{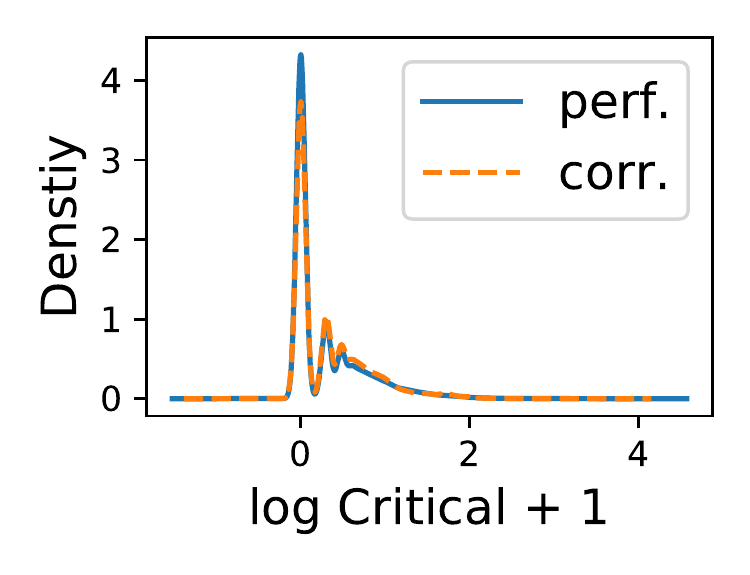}
    \caption{Density plot of metric values for perfective and corrective categories before the change.}
    \label{fig:density_parent}
\end{figure}
Figure~\ref{fig:density_parent} shows another perspective on our data in the form of a direct comparison of the density between perfective and corrective changes.
We can see that McCC, NLE, LLOC, NUMPAR, CD, CBO, NII and Minor have a lower density for perfective than for corrective.
While the differences are small they are noticeable.

\begin{center}
    \setlength{\fboxrule}{0pt}
        \vrule\vrule\vrule\vrule\vrule\vrule\fcolorbox{white}{gray!15}{
            \hspace{.2em}\parbox{.91\linewidth} {
                \vspace{.5em}
                \textbf{RQ2 Summary}\\
                The files that are targets of perfective changes are in median not large and complex even before the change is applied.
                Corrective changes are applied to files which are in median already complex and large.
                The differences are statistically significant for most metrics, however the effect sizes are negligible to small.
                \vspace{.5em}
        }\hspace{.5em}
    }
\end{center}

\section{Discussion}\label{sec:discussion}

Our results show that size is different in both types of commits in \textbf{H1}.
The size difference between all commits and perfective as well as corrective commits shows that both tend to be smaller than other commits.
In case of perfective commits, code is statistically significantly more often deleted.

The differences in change size\del{s} as well as the increased number of deletions for perfective commits we found for \textbf{H1} confirms previous research.
\del{Multiple studies, e.g.,}\rev{The studies by }\cite{Mockus2000}, \cite{Purushothaman2005} and \cite{Alali2008} found that perfective maintenance activities are usually smaller.
\cite{Mockus2000} as well as \cite{Purushothaman2005} found that corrective maintenance is also smaller and that perfective maintenance deletes more code.
Another indication that size between maintenance types is different can be seen in the work by \cite{Honel2019}, which used size based metrics as predictors for maintenance types and showed that it improved the performance
of classification models.

Our results for \textbf{H2} show statistically significant differences in metric measurements between perfective commits and other commits.
This result indicates a confirmation of the measurements used by quality models, as the majority of metrics change as expected when developers actively improve the internal code quality.
This empirical confirmation of the connection between quality metrics and developer intent is one of our main contributions and was, to the best of our knowledge, not part of any prior study.
However, there are several examples of prior work that assumed this relationship.

The publications by \cite{mccabe} and \cite{ckmetrics} assume that reducing complexity and coupling metrics increases software quality which is in line with our developer intents.
While all metrics are included in a current ColumbusQM version~\citep{Bakota2014} because we used it as a basis, the CBO, McCC, LLOC, NOA metrics are also part of the SQUALE model~\citep{squale}
AD, NLE, McCC, and PMD warnings are also part of Quamoco~\citep{Wagner2012}.
It seems that developers and the Columbus quality model agree with their view on software quality. We find that most of the metrics used in the quality model change when developers perceive their change as quality increasing.
This is also true for most of the metrics shared with the SQUALE model and with the Quamoco quality model.
However, the implementation for the metrics may differ between the models.
Our work establishes that all these quality models are directly related to intended improvements of the internal code quality by the developers.

Surprisingly, we found only few statistically significant and non\rev{-}negligible differences for corrective commits.
Not all software metric values are changing into the expected direction for corrective commits.
For example, we can see that McCC, LLOC and NLE are increasing in corrective changes compared to other commits.
While we are not expecting them to decrease for every corrective commit, we assumed that in comparison to all other commits they would be decreasing.
Even when considering software aging~\citep{Parnas2001} we would expect the aging to impact all kinds of changes not just corrective changes.
When we look at popular data sets used in the defect prediction domain we often find coupling, size and complexity software metrics~\citep{Datensatz}.
For example, the popular (as per the literature review from \cite{Hosseini2017}) data set by \cite{Jureczko2010} uses such features, but they are also common in more recent data sets, e.g., by \cite{Ferenc2020} or \cite{Yatish2019}.

That the most significant difference is in the size of changes could explain various recent findings from the literature, in which size was found to be a very good indicator both for release level defect prediction~\citep{Zhou2018} and just-in-time defect prediction~\citep{Huang2017}. This could also be an explanation for possible ceiling effects~\citep{Ceiling} when such criteria are used, as the difference to other changes are relatively small. We believe that these aspects should be further considered by the defect prediction community and believe that more research is required to establish causal relationships between features and defectiveness.

While the work by \cite{fmri} indicates that cyclomatic complexity may not be as indicative of code understandability as expected, we show within our work that it often changes in quality increasing commits.
It seems that developers associate overall complexity as measured by McCC, NLE, NUMPAR with code that needs quality improvement. However, as we can see in the exploratory part of our study the most complex files are usually not targeted for quality increasing changes.

Our exploratory study to answer \textbf{RQ2} about files that are the target of quality increasing commits reveals additional interesting data.
We show that perfective maintenance does not necessarily target files that are in need of it due to high complexity in comparison to other changes.
In fact, low complexity files as measured by McCC and NLE are more often part of additional quality increasing work by the developers.
This may hint at problems regarding the prioritization of quality improvements in the source code. Maybe errors could have been avoided when perfective changes would have targeted more complex files.
There could also be effects of different developers or a bias for perfective changes towards simpler code, this warrants future investigation.
Corrective changes, in contrast to perfective changes, are applied to files which are large and complex. This was expected, however combined with the results of \textbf{RQ1} this means that bugs are fixed in complex and large files and then the files get, on average, even more complex and even larger.

Future work could investigate boundary values according to our data.
When we compare the median values of our measurements in Table~\ref{tbl:median} with current boundary values from PMD\footnote{https://pmd.github.io/pmd/pmd\_rules\_java\_design.html\#cyclomaticcomplexity}, we may think that the PMD warning value of 80 McCC per file may be too high.
A PMD warning triggered at 34 McCC per file would have warned about at least 50\% of the files that were in need of a bug fix.
However, lowering the boundary will also result in more warnings for files that were not target of corrective changes.

\subsection{\rev{Implications for Researchers}}
\rev{Our results for \textbf{H1} increase the validity of previous research by confirming previous results in our study on a larger data set of different projects.
Our confirmation that quality increasing changes are smaller than other changes shows that researchers developing a change classification approach can benefit from including size based metrics.}

\rev{Our results for \textbf{H2} show that perfective changes reduce size and complexity metrics in comparison to all other changes.
Previous studies investigating refactorings also found an impact on size and complexity metrics.
We are able to generalize this finding by providing results of a superset of refactoring operations, namely perfective changes.
This indicates that perfective changes generally reduce size and complexity metrics.
This also indicates that software quality models that use the affected metrics in their code quality estimations agree with the developers on what impacts code quality.}

\rev{Increasing the external quality by fixing bugs, i.e., corrective changes, decreases the internal quality, i.e., complexity metric values.
Defect prediction models may assign a higher risk to parts of the code that contained a bug before as there is an assumption of latent bugs still existing~\citep{fixcache,bugcache}.
Our data provides a fine grained perspective by providing empirical data which shows that the code quality as measured by static source code metrics is actually decreasing.}

\rev{This also has implications for researchers developing and deploying defect prediction models in practice.
The fact that fixing a bug increases the risk of the file can lead to problems regarding the acceptance of the model by practitioners as they have no way of reducing the risk~\citep{Lewis2013}.
The results of our study could help to explain the reasons to developers.
We can empirically show that fixing a bug is a complex operation that introduces even more complexity than other changes, even feature additions.
According to our results, the main driver of complexity in a project are bug fixes and the only way to combat the rising complexity is perfective maintenance which should especially target large and complex files.}

\rev{In our results for \textbf{RQ2} we see a difference between files before corrective changes are applied\rev{,} and before other changes are applied.
This difference is one of the sources of the predictive power of defect prediction models. However, the difference is smaller than expected.
Incorporating metrics that have a larger difference in our data, e.g., comment density and API documentation into defect prediction models, may increase their prediction performance.}

\subsection{\rev{Implications for Practitioners}}

\rev{Our results for \textbf{H2} suggest that, for the most part, software quality models match the expectations of the developers.
If practitioners select a software quality model which uses static source code metrics that show a difference in our data they can expect that the model matches their intuition.}

\rev{In combination with \textbf{RQ2}, our results indicate that bug fixing is the main driver of complexity in a software project and perfective changes are the main reducer of complexity.
This has implications for developers. If more complex files were targeted for perfective maintenance bugs could possibly have been prevented.
As fixing bugs does not decrease complexity, perfective maintenance is the best way to reduce it and combat rising complexity of the project as a whole.
However, given the results for \textbf{RQ2}, we see that large and complex files are not the main target of perfective maintenance.
This is an opportunity for improvement by shifting priorities for perfective maintenance to large and complex files.
Moreover, our results indicate that a bug fix should be treated similar to technical debt regarding its negative impact on complexity metrics.
To mitigate this, practitioners should be aware that it would be beneficial to clean up and simplify the code that is introduced as part of the bug fix.}

\section{Threats to validity}
\label{sec:threats_to_validity}

In this section, we discuss the threats to validity we identified for our work.
We discuss four basic types of validity separately as suggested by \cite{Wohlin} and include reliability due to our manual classification approach.

\subsection{Reliability}
We classify changes to a software retroactively and without the developers. This may introduce a researcher bias to the data and subsequently the results.
However, this is a necessity given the size of the data and the unrestricted time frame for the sample and full data because it would not be feasible to ask developers about a couple of commits from years ago.
To mitigate this threat, we perform the classification labeling according to guidelines and every change is independently classified by two researchers.
We also compare our differences with a sample of changes classified by the developers themselves from \cite{Mauczka2015} and confirm that we are agreeing on most changes.
In addition, we measure the inter-rater agreement between the researchers and find that it is substantial.

\subsection{Construct Validity}
Our definition of quality improving may be too broad. We aggregate different types of quality improvement together, e.g., improving error messages, structure of the code or readability.
This may influence the changes we observe within our metric values. While these differences should be studied as well, we believe that a broad overview of generic quality improvements independent of their type has advantages. We avoid the risk of being focused only on structural improvements, i.e., due to use of generics or new Java features without missing bigger changes due to simplification of method code.

\subsection{Conclusion Validity}
We are reporting differences in metric value changes between perfective and corrective changes of the software development history of our study subjects.
We find a difference for perfective commits and only some non\rev{-}negligible, statistically significant difference for corrective commits.
This could be an effect of our sample used as ground truth, however we chose to draw randomly from a list of commits in our study subjects so that our sample should be representative.

We use a deep learning model to classify all of our commits based on the ground truth we provide. This can introduce a bias or errors in the classification. We note however, that the non-negligible effect sizes for our results do not change.
The quality metric evaluation of only the ground truth data is included in the appendix and shows similar results.
We note that for the small effect sizes we observe, a large number of observations are needed to show a significant difference as is demonstrated by the results in this article when compared to the ground truth.

\subsection{Internal Validity}
A possible threat could be tangled commits which improve quality and at the same time add a feature.
We mitigate this in our ground truth, by manual inspection of the commit message of every change considered. We excluded tangled commits if it was possible to determine this by the commit message.
As no automatic untangling approach is available to us and available approaches to label tangled commits already use the commit message to find tangled commits we determine that tangled commits which are not identifiable from the commit message are a minor threat.

Another threat could be a lower number of feature additions in our study subjects. Maybe feature additions happen too infrequently to influence the results, therefore, corrective commits are seen as adding more complex code than other commits.
While we include some projects that are in development for a long period of time, we believe this threat is mitigated by the unrestricted time frame of our study.

\rev{Bots which commit code~\citep{Dey2020} could be a possible threat to our study.
We mitigate this threat by matching our author data against the bot data set provided by~\cite{Dey2020}. We did not find matches for bots in our data.
We were able to detect a Jenkins bot only when dropping the restriction of our case study data that a commit has to change non-test code.
We also implemented the detection mechanism by~\cite{Dey2020} which uses the username and email of the author of the commit, as used by Dey et al. to create their bot data set.
This also yielded no bots in our data. Manual inspection of the author data yielded two bot-like accounts which turned out to be from a previous cvs2svn conversion as well as asf-sync-process which allows user patches without an account.
However, the content of changes by the accounts we found are created by developers. We determine that the threat of bots in our data is low.}

\subsection{External Validity}
We focus on a convenience sample of data consisting of Java Open Source projects under the umbrella of the Apache Software Foundation.
We consider this a minor threat to external validity. The reason is that although we are limited to one organization, we still have a wide variety of different types of software in our data. We believe that this mitigates the missing variety of project patronage.

Furthermore, we only include Java projects. However, Java is used in a wide variety of projects and remains a popular language. Its age provides us with a long history of data we can utilize in this study.
However, we note that this study may not generalize to all Java projects much less all software projects in other languages.

\section{Conclusion}
\label{sec:conclusion}

Numerous quality measurements exist, and numerous software quality models try to connect concrete quality metrics with abstract quality factors and sub factors.
Although it seems clear that some static source code metrics influence software quality factors, the question of which and how much remains.
Instead of relying on necessarily limited developer and expert evaluations of source code or changes we extract metrics from past changes where developers intended to increase the quality extracted from the commit message.

Within this work, we performed a manual classification of developer intents on a sample of \numberCommits{} commits from \numberProjects{} Java open source projects by two researchers independently and guided by classification guidelines.
We classify the commits into three categories, perfective maintenance, corrective maintenance, or neither. We further evaluate our classification guidelines by re-classifying of a developer labeled sample.
We use the manually labeled data as ground truth to evaluate and then fine tune a state-of-the-art deep learning model for text classification. The fine-tuned model is then used to classify all available commits into our categories increasing our data size to \numberCommitsAll{} commits.
We extract static source code metrics and static analysis warnings for all \numberCommitsAll{} commits which allows us to investigate the impact of changes and the distribution of metric values before the changes are applied.
Based on the literature, we hypothesize that certain metric values change in a certain direction, e.g., perfective changes reduce complexity.
We find that perfective commits are more often removing code and generally add \del{less}\rev{fewer} lines.
Regarding the metric measurements, we find that most metric value changes of perfective commits are significantly different to other commits and have a positive, non\rev{-}negligible impact on the majority of metric values.

Surprisingly, we found that corrective changes are more complex and larger than other changes. It seems that fixing a bug increases the size\rev{,} but also the complexity measured via McCC and NLE.
As we compare against all other changes, we were expecting less addition of complexity as e.g., feature additions.
We conclude that the process of performing a bug fix tends to add more complex code than other changes.

We find that complex files are not necessarily the primary target for quality increasing work by developers, including refactoring.
To the contrary, we find that perfective quality changes are applied to files that are already less complex than files changed in other or corrective commits.
Files contained in corrective changes on the other hand are more complex and usually larger than both perfective and other files. In combination with our first result this shows that corrective changes are applied to files which are already complex and get even more complex after the change is applied.

While we explored a limited number of metrics and commits we think that this approach can be used to evaluate more metrics connected with software quality in a meaningful way and help practitioners and researchers with additional empirical data.

\section*{Declarations}
This work was partly funded by the German Research Foundation (DFG) through the project DEFECTS, grant 402774445.\\
The authors have no competing interests to declare that are relevant to the content of this article.

\section*{Acknowledgements}
We want to thank the GWDG Göttingen\footnote{https://www.gwdg.de} for providing us with computing resources within their HPC-Cluster.

\bibliographystyle{spbasic}
\bibliography{literature}

\clearpage
\begin{appendices}
    \section{Ground truth only results}

\begin{figure}[h]
    \includegraphics[width=0.25\textwidth]{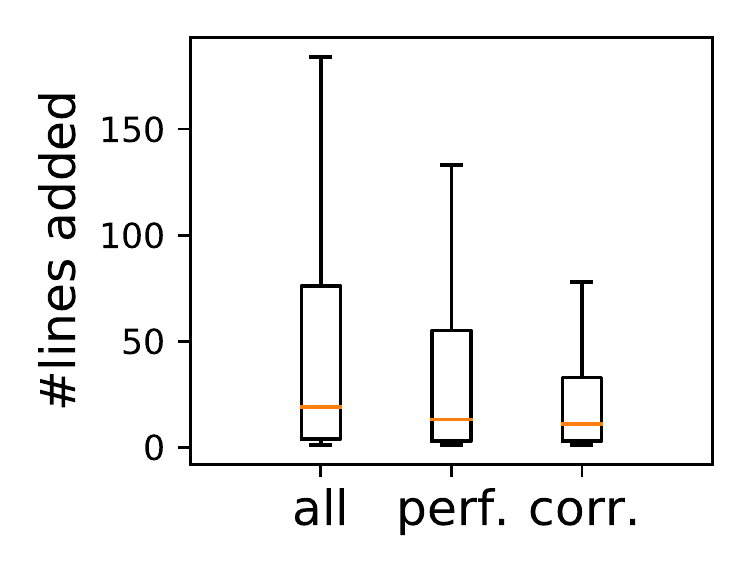}\includegraphics[width=0.25\textwidth]{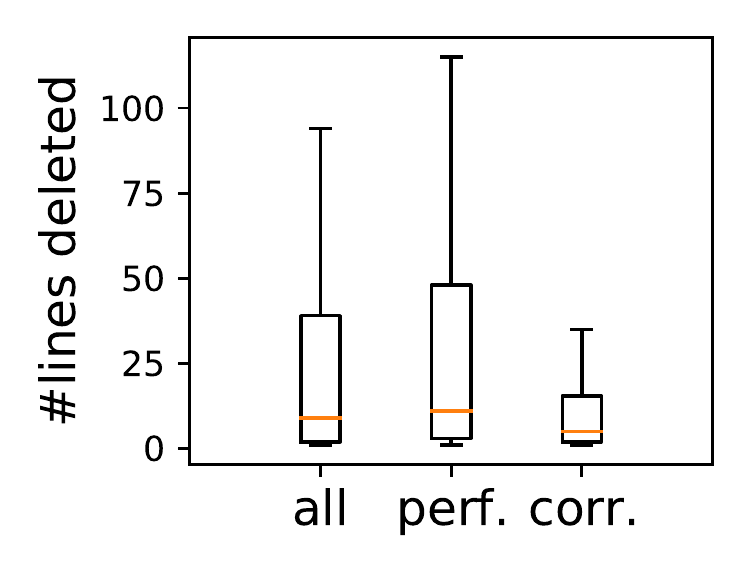}
    \includegraphics[width=0.25\textwidth]{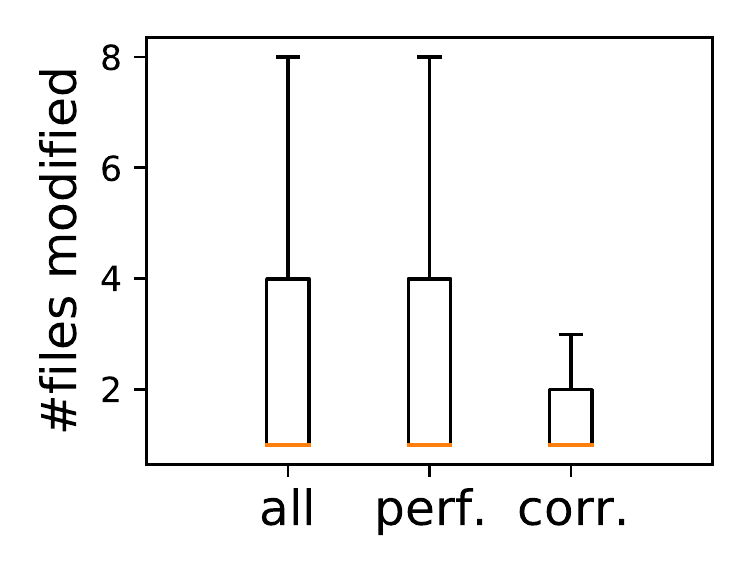}\includegraphics[width=0.25\textwidth]{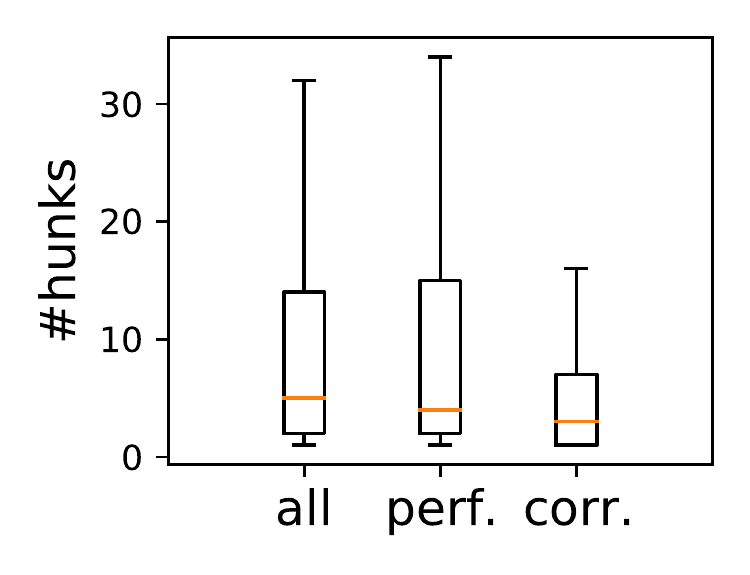}
    \caption{Ground truth only. Commit size distribution over all projects for all, perfective and corrective commits. Fliers are omitted.}
    \label{fig:bp_dist_gt}
\end{figure}

\begin{table}[h]
    \caption{Ground truth only. Statistical test results for perfective and corrective commits, Mann-Whitney U test p-values ($p$-value) and effect size ($d$) with category $n$ is negligible, $s$ is small. Statistically significant $p$-values are bolded.}\label{tbl:rq1_stats_gt}
    \centering
    \begin{tabular}{lrr|rr}
        & \multicolumn{2}{c}{Perfective} & \multicolumn{2}{c}{Corrective}\\
        \toprule
        Metric & $p$-value & $d$ & $p$-value & $d$\\
        \midrule
\#lines added & \textbf{\textless0.0001} & 0.20 (s) & \textbf{\textless0.0001} & 0.20 (s)\\
\#lines deleted & \textbf{\textless0.0001} & 0.13 (s) & \textbf{\textless0.0001} & 0.17 (s)\\
\#files modified & 0.2829 & - & \textbf{\textless0.0001} & 0.22 (s)\\
\#hunks & 0.7009 & - & \textbf{\textless0.0001} & 0.21 (s)\\
        \bottomrule
    \end{tabular}
\end{table}

\begin{figure}[h]
    \includegraphics[width=0.24\textwidth]{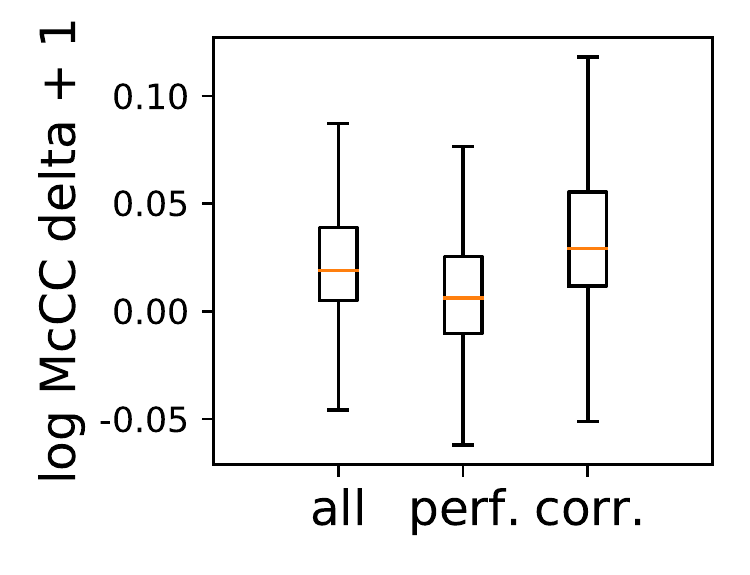}\includegraphics[width=0.24\textwidth]{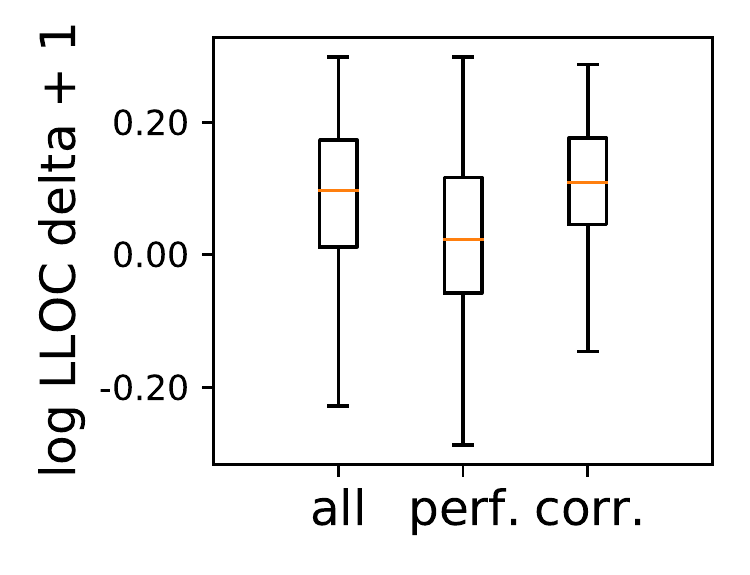}
    \includegraphics[width=0.24\textwidth]{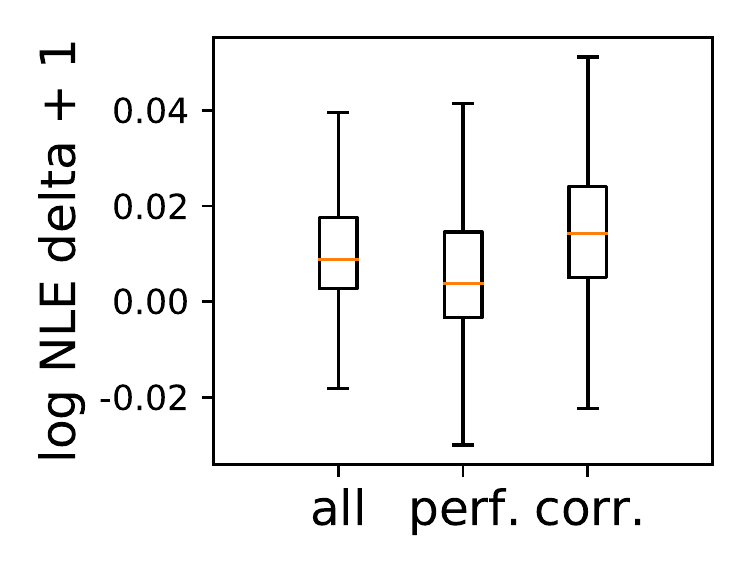}\includegraphics[width=0.24\textwidth]{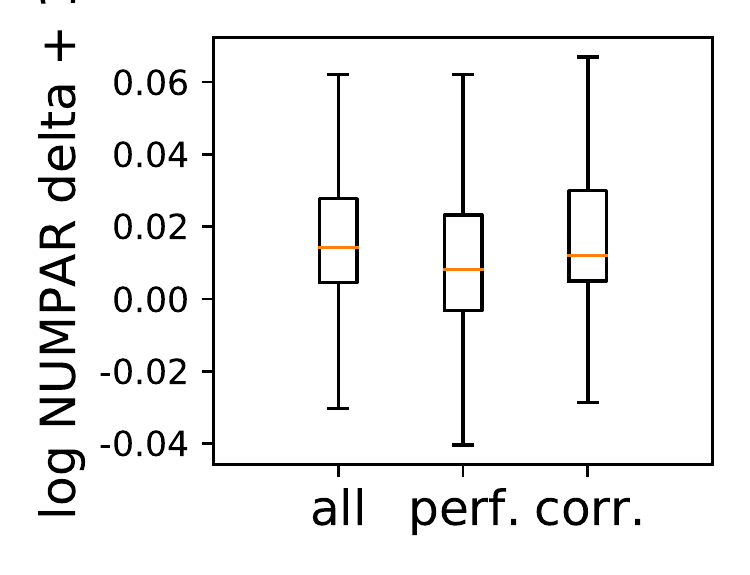}
    \includegraphics[width=0.24\textwidth]{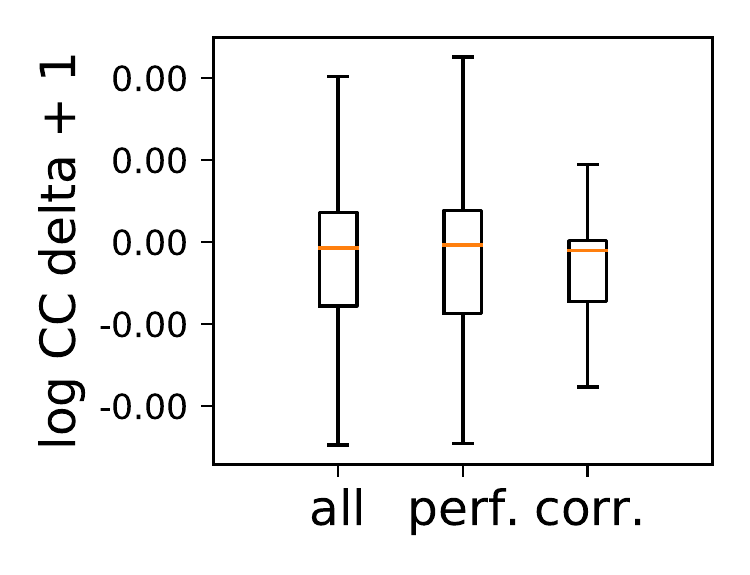}\includegraphics[width=0.24\textwidth]{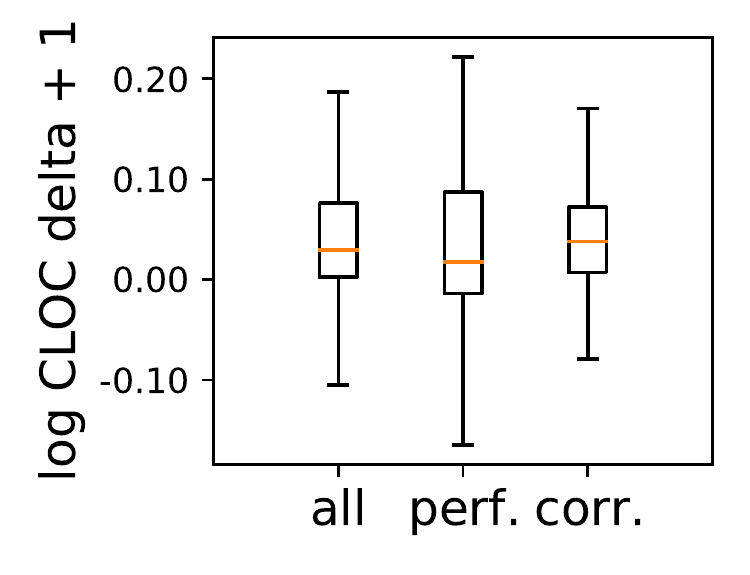}
    \includegraphics[width=0.24\textwidth]{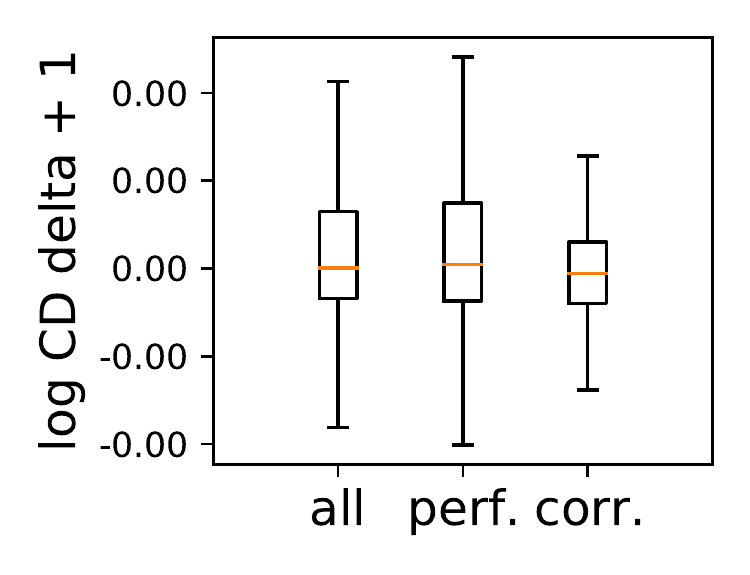}\includegraphics[width=0.24\textwidth]{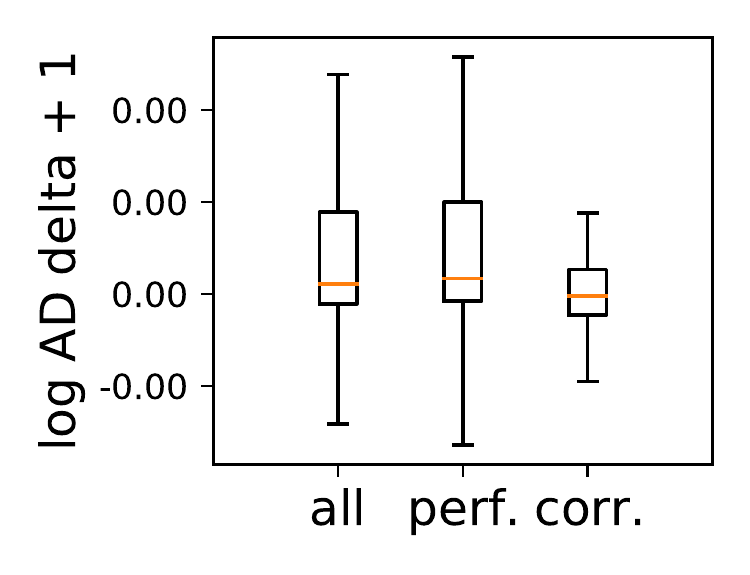}
    \includegraphics[width=0.24\textwidth]{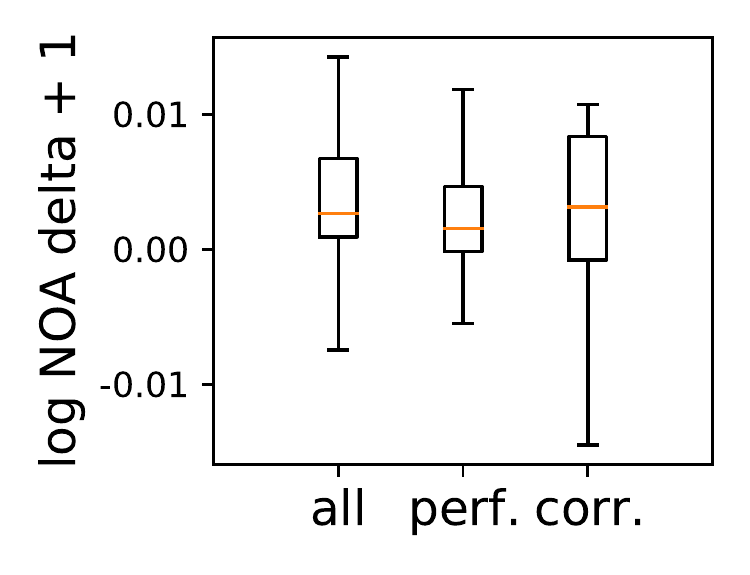}\includegraphics[width=0.24\textwidth]{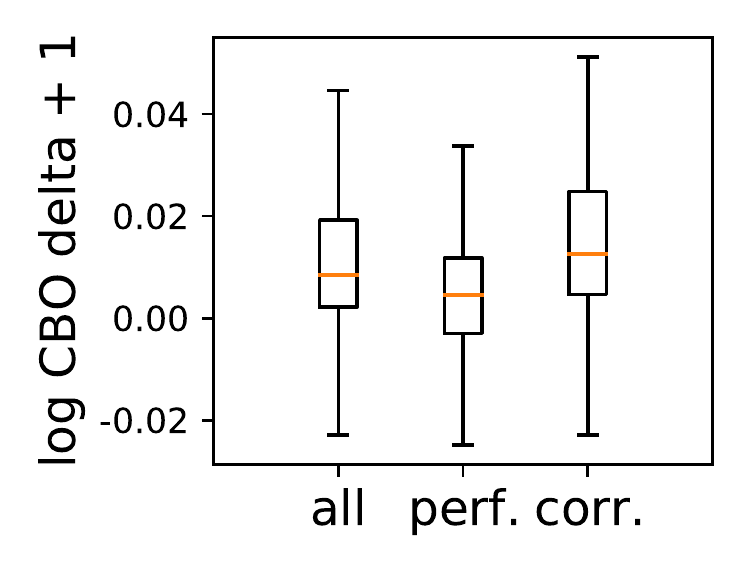}
    \includegraphics[width=0.24\textwidth]{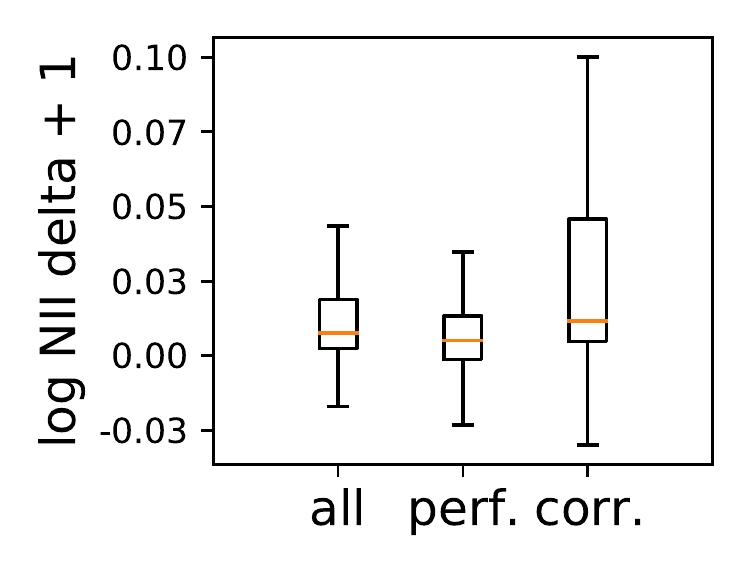}\includegraphics[width=0.24\textwidth]{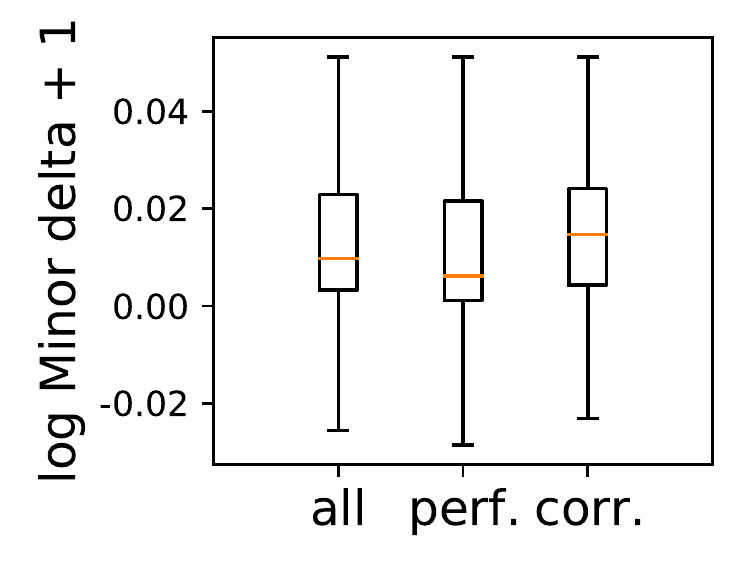}
    \includegraphics[width=0.24\textwidth]{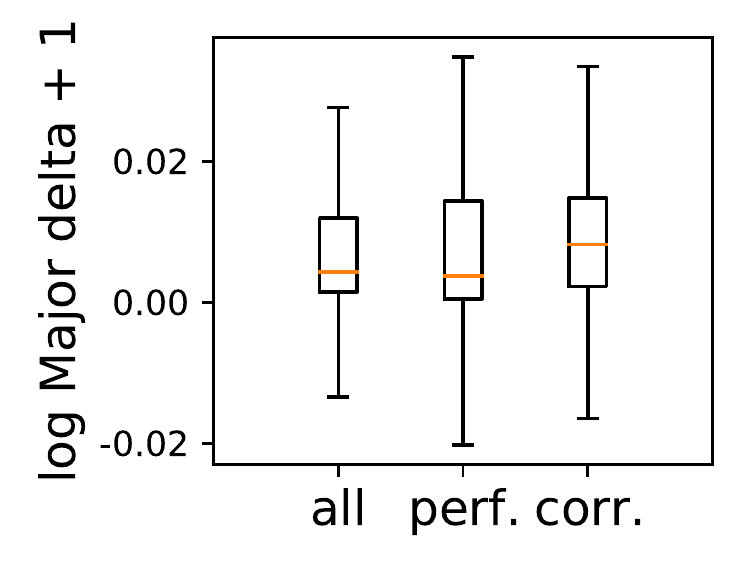}\includegraphics[width=0.24\textwidth]{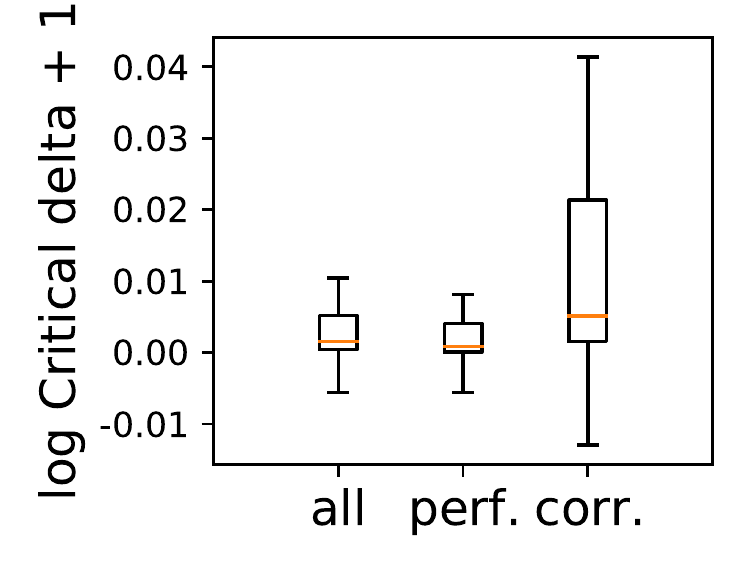}
    \caption{Ground truth only. Static source code metrics changes in all, perfective and corrective commits divided by changed lines. Fliers are omitted.}
    \label{fig:bp_static_density_gt}
\end{figure}

\begin{table}[h]
    \caption{Ground truth only. Statistical test results for perfective and corrective commits, Mann-Whitney U test p-values ($p$-value) and effect size ($d$) with category, $n$ is negligible, $s$ is small, $m$ is medium. Statistically significant $p$-values are bolded. All values are normalized for changed lines.}\label{tbl:diff_lloc_gt}
    \centering
    \begin{tabular}{lrr|rr}
        & \multicolumn{2}{c}{Perfective} & \multicolumn{2}{c}{Corrective}\\
        \toprule
        Metric & $p$-val & $d$ & $p$-val & $d$\\
McCC & \textbf{\textless0.0001} & 0.37 (m) & 1.0000 & -\\
LLOC & \textbf{\textless0.0001} & 0.42 (m) & 1.0000 & -\\
NLE & \textbf{\textless0.0001} & 0.26 (s) & 0.9577 & -\\
NUMPAR & \textbf{\textless0.0001} & 0.24 (s) & \textbf{\textless0.0001} & 0.09 (n)\\
CC & 1.0000 & - & \textbf{\textless0.0001} & 0.12 (s)\\
CLOC & \textbf{\textless0.0001} & 0.19 (s) & 0.1906 & -\\
CD & 0.9303 & - & \textbf{\textless0.0001} & 0.15 (s)\\
AD & 0.1556 & - & \textbf{\textless0.0001} & 0.10 (s)\\
NOA & \textbf{\textless0.0001} & 0.08 (n) & \textbf{\textless0.0001} & 0.09 (n)\\
CBO & \textbf{\textless0.0001} & 0.18 (s) & 0.0145 & -\\
NII & \textbf{\textless0.0001} & 0.19 (s) & 0.0620 & -\\
Minor & \textbf{\textless0.0001} & 0.18 (s) & 0.0005 & -\\
Major & \textbf{\textless0.0001} & 0.10 (s) & \textbf{0.0002} & 0.06 (n)\\
Critical & \textbf{\textless0.0001} & 0.06 (n) & 0.1111 & -\\
        \midrule
        \bottomrule
    \end{tabular}
\end{table}

\begin{figure}[h]
    \includegraphics[width=0.24\textwidth]{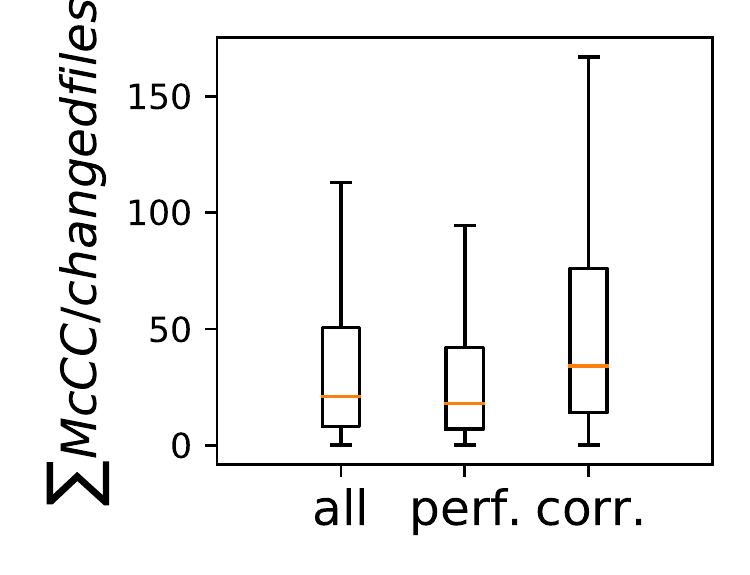}\includegraphics[width=0.24\textwidth]{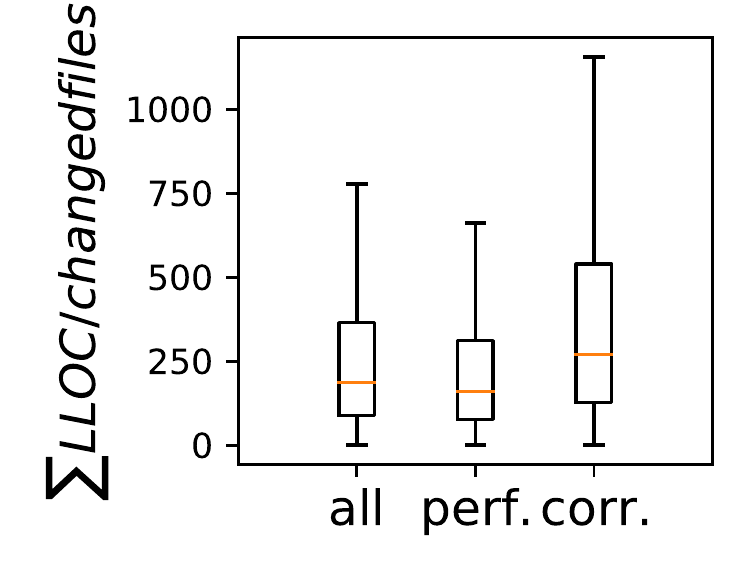}
    \includegraphics[width=0.24\textwidth]{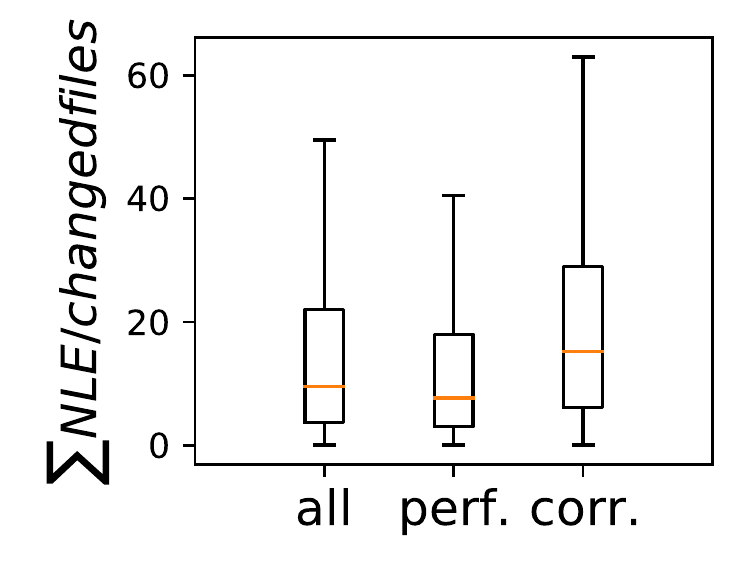}\includegraphics[width=0.24\textwidth]{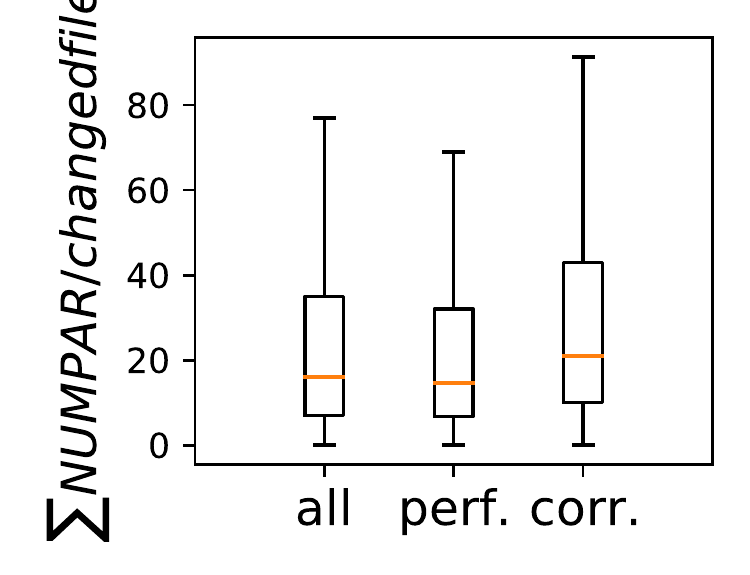}
    \includegraphics[width=0.24\textwidth]{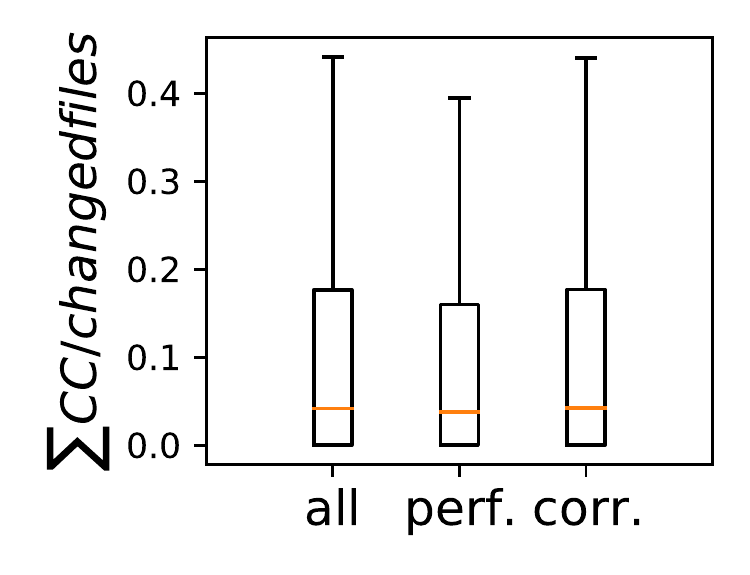}\includegraphics[width=0.24\textwidth]{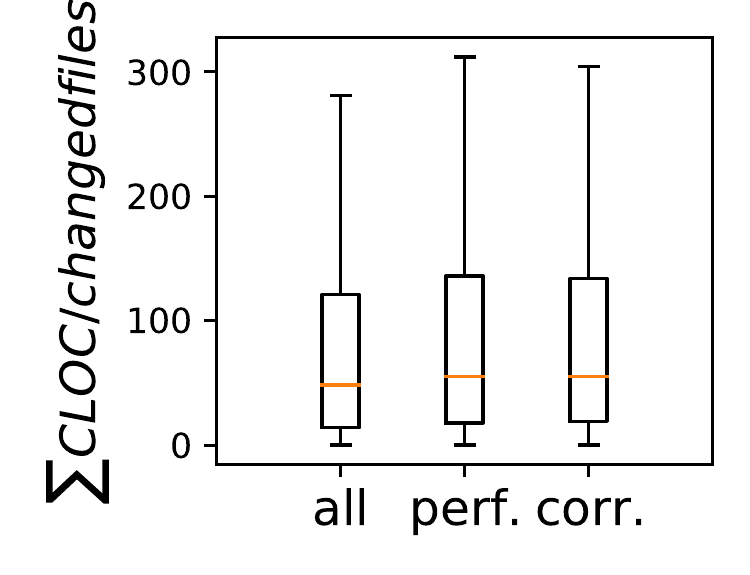}
    \includegraphics[width=0.24\textwidth]{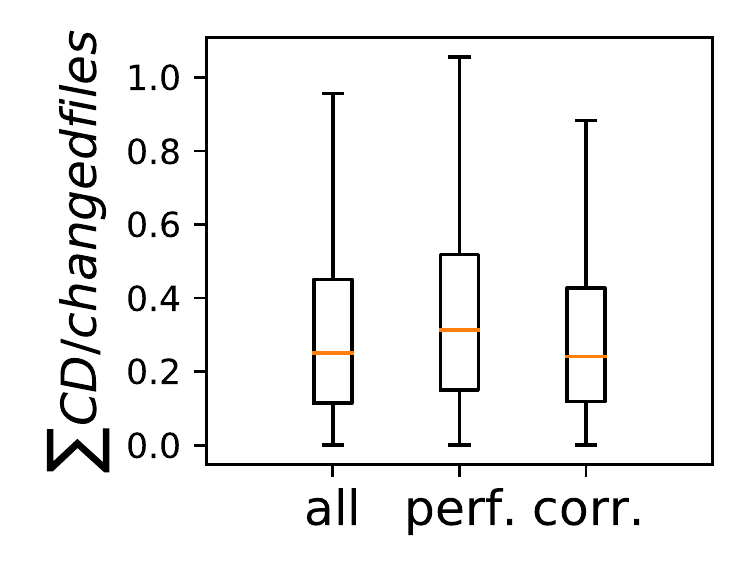}\includegraphics[width=0.24\textwidth]{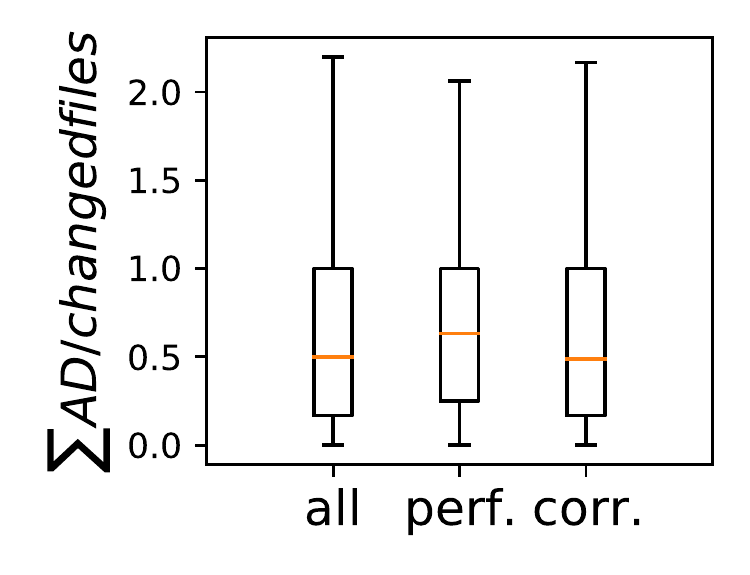}
    \includegraphics[width=0.24\textwidth]{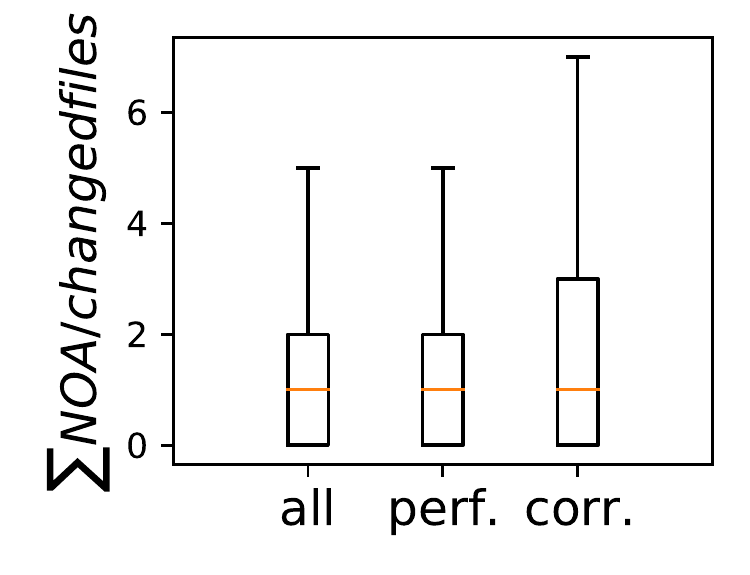}\includegraphics[width=0.24\textwidth]{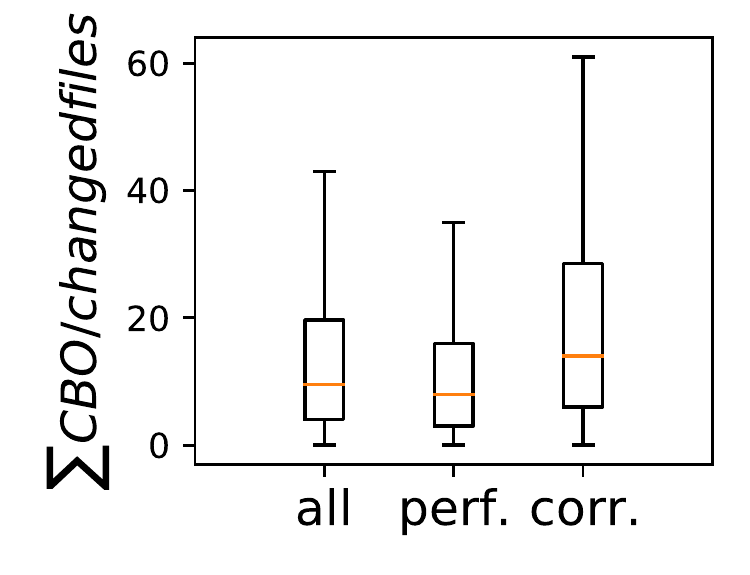}
    \includegraphics[width=0.24\textwidth]{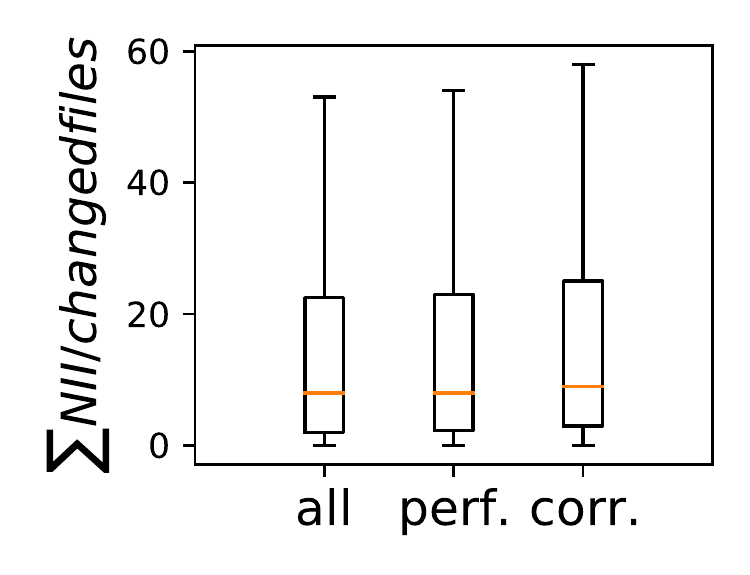}\includegraphics[width=0.24\textwidth]{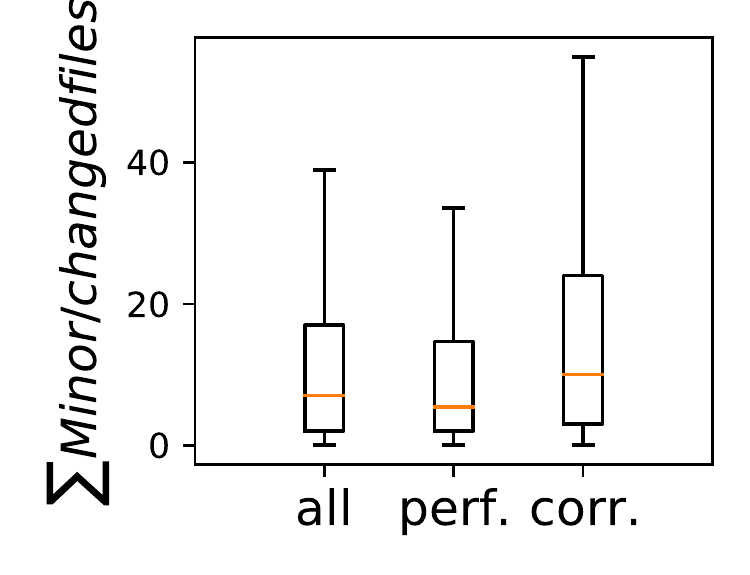}
    \includegraphics[width=0.24\textwidth]{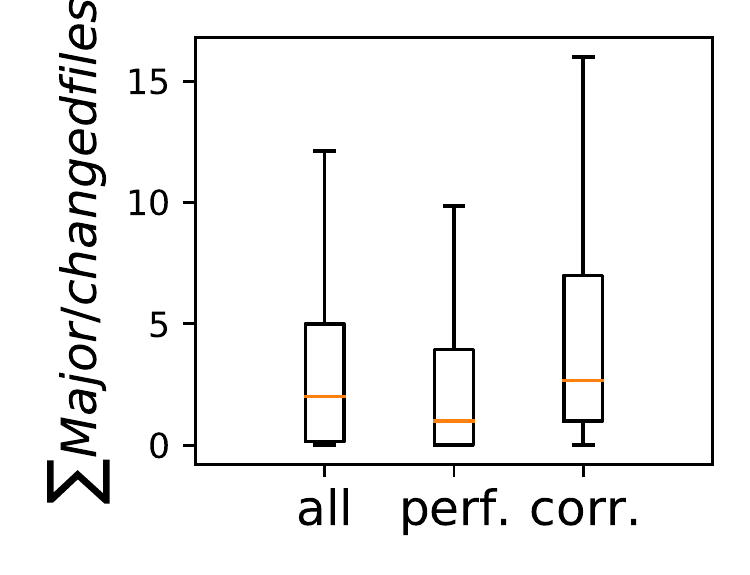}\includegraphics[width=0.24\textwidth]{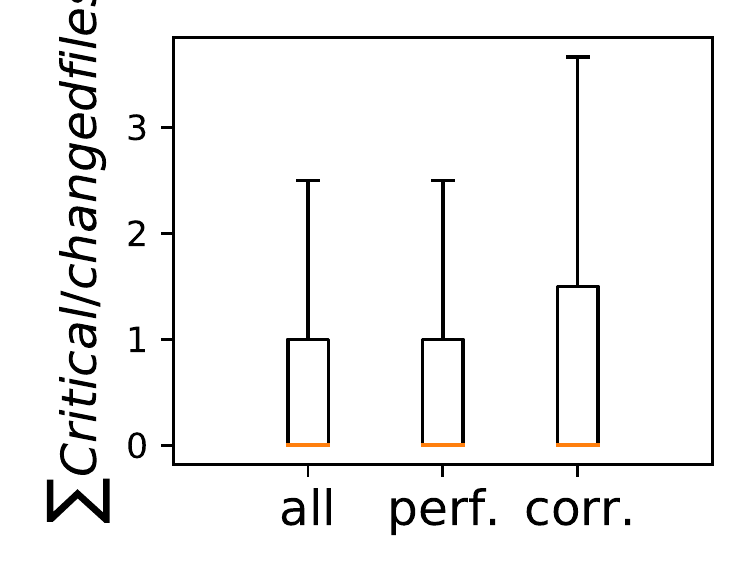}
    \caption{Ground truth only. Static source code metrics before the change is applied. Fliers are omitted.}
    \label{fig:bp_static_parent_gt}
\end{figure}

\begin{table}[h]
    \caption{Median metric values before the change is applied.}\label{tbl:median_gt}
    \centering
    \begin{tabular}{lrrr}
        \toprule
        Metric & All & Perfective & Corrective\\
        \midrule
McCC & 21.00 & 18.00 & 34.00\\
LLOC & 187.22 & 160.38 & 270.00\\
NLE & 9.50 & 7.67 & 15.20\\
NUMPAR & 16.00 & 14.67 & 21.00\\
CC & 0.04 & 0.04 & 0.04\\
CLOC & 48.22 & 55.00 & 55.00\\
CD & 0.25 & 0.31 & 0.24\\
AD & 0.50 & 0.63 & 0.49\\
NOA & 1.00 & 1.00 & 1.00\\
CBO & 9.50 & 8.00 & 14.00\\
NII & 8.00 & 8.00 & 9.00\\
Minor & 7.00 & 5.43 & 10.00\\
Major & 2.00 & 1.00 & 2.67\\
Critical & 0.00 & 0.00 & 0.00\\
\bottomrule
    \end{tabular}
\end{table}

\begin{table}[h]
    \caption{Ground truth only. Statistical test results for perfective and corrective commits regarding their average metrics before the change, Mann-Whitney U test p-values ($p$-value) and effect size ($d$) with category, $n$ is negligible, $s$ is small, $m$ is medium. Statistically significant $p$-values are bolded.}\label{tbl:stats_parent_gt}
    \centering
    \begin{tabular}{lrr|rr}
        & \multicolumn{2}{c}{Perfective} & \multicolumn{2}{c}{Corrective}\\
        \toprule
        Metric & $p$-val & $d$ & $p$-val & $d$\\
        \midrule
McCC & 0.0003 & - & 0.0016 & -\\
LLOC & 0.0005 & - & 0.1138 & -\\
NLE & 0.0003 & - & 0.0072 & -\\
NUMPAR & 0.5344 & - & 0.4704 & -\\
CC & 0.4142 & - & 0.0210 & -\\
CLOC & \textbf{\textless0.0001} & 0.10 (n) & 0.0111 & -\\
CD & \textbf{\textless0.0001} & 0.15 (s) & \textbf{\textless0.0001} & 0.16 (s)\\
AD & \textbf{\textless0.0001} & 0.15 (s) & \textbf{\textless0.0001} & 0.15 (s)\\
NOA & 0.6847 & - & 0.2103 & -\\
CBO & \textbf{\textless0.0001} & 0.11 (s) & 0.0190 & -\\
NII & 0.0510 & - & 0.0105 & -\\
Minor & 0.0006 & - & 0.6288 & -\\
Major & \textbf{\textless0.0001} & 0.12 (s) & 0.0852 & -\\
Critical & 0.0179 & - & 0.5730 & -\\
    \bottomrule
    \end{tabular}
\end{table}

\begin{table}[h]
    \caption{\rev{Detailed statistical tests results for metric changes. Accompanies Table~\ref{tbl:diff_lloc}.
    SHA is Shapiro-Wilk, MWU is Mann-Whitney U test, the number of samples for not perfective is 77,630, for perfective the number is 47,852.
    The number of samples for not corrective is 90,258 and for corrective 35,124.
For both samples the median, Shapiro-Wilk test statistic and p-value are given comma separated.}}
    \centering
    \begin{tabular}{lrrrr}
    \multicolumn{5}{c}{Perfective changes}\\
    \toprule
Metric & MWU Statistic & Median & SHA Statistic & SHA p-val\\
    \midrule
McCC & 2579401012.5 & 0.02,0.00 & 0.55,0.27 & \textbf{\textless0.0001},\textbf{\textless0.0001}\\
LLOC & 2691844899.5 & 0.25,0.00 & 0.57,0.20 & \textbf{\textless0.0001},\textbf{\textless0.0001}\\
NLE & 2351847133.0 & 0.00,0.00 & 0.58,0.20 & \textbf{\textless0.0001},\textbf{\textless0.0001}\\
NUMPAR & 2328626543.0 & 0.00,0.00 & 0.39,0.05 & \textbf{\textless0.0001},\textbf{\textless0.0001}\\
CC & 1666541612.5 & 0.00,0.00 & 0.03,0.01 & \textbf{\textless0.0001},\textbf{\textless0.0001}\\
CLOC & 2158356261.5 & 0.00,0.00 & 0.32,0.34 & \textbf{\textless0.0001},\textbf{\textless0.0001}\\
CD & 1715608163.5 & 0.00,0.00 & 0.41,0.21 & \textbf{\textless0.0001},\textbf{\textless0.0001}\\
AD & 1899339427.0 & 0.00,0.00 & 0.25,0.13 & \textbf{\textless0.0001},\textbf{\textless0.0001}\\
NOA & 1997259809.5 & 0.00,0.00 & 0.06,0.01 & \textbf{\textless0.0001},\textbf{\textless0.0001}\\
CBO & 2208901912.0 & 0.00,0.00 & 0.21,0.04 & \textbf{\textless0.0001},\textbf{\textless0.0001}\\
NII & 2210463268.0 & 0.00,0.00 & 0.09,0.07 & \textbf{\textless0.0001},\textbf{\textless0.0001}\\
Minor & 2201853734.0 & 0.00,0.00 & 0.04,0.01 & \textbf{\textless0.0001},\textbf{\textless0.0001}\\
Major & 2077680338.5 & 0.00,0.00 & 0.04,0.00 & \textbf{\textless0.0001},\textbf{\textless0.0001}\\
Critical & 1952568002.5 & 0.00,0.00 & 0.05,0.05 & \textbf{\textless0.0001},\textbf{\textless0.0001}\\
    \midrule
    \multicolumn{5}{c}{Corrective changes}\\
    \midrule
McCC & 1319862052.5 & 0.00,0.00 & 0.36,0.36 & \textbf{\textless0.0001},\textbf{\textless0.0001}\\
LLOC & 1406100592.5 & 0.07,0.18 & 0.36,0.36 & \textbf{\textless0.0001},\textbf{\textless0.0001}\\
NLE & 1538986445.5 & 0.00,0.00 & 0.35,0.35 & \textbf{\textless0.0001},\textbf{\textless0.0001}\\
NUMPAR & 1736495605.5 & 0.00,0.00 & 0.14,0.14 & \textbf{\textless0.0001},\textbf{\textless0.0001}\\
CC & 1781604826.0 & 0.00,0.00 & 0.01,0.01 & \textbf{\textless0.0001},\textbf{\textless0.0001}\\
CLOC & 1665288104.5 & 0.00,0.00 & 0.38,0.38 & \textbf{\textless0.0001},\textbf{\textless0.0001}\\
CD & 1833218654.0 & 0.00,0.00 & 0.28,0.28 & \textbf{\textless0.0001},\textbf{\textless0.0001}\\
AD & 1719709796.5 & 0.00,0.00 & 0.19,0.19 & \textbf{\textless0.0001},\textbf{\textless0.0001}\\
NOA & 1700427713.0 & 0.00,0.00 & 0.03,0.03 & \textbf{\textless0.0001},\textbf{\textless0.0001}\\
CBO & 1687001103.5 & 0.00,0.00 & 0.09,0.09 & \textbf{\textless0.0001},\textbf{\textless0.0001}\\
NII & 1621472694.0 & 0.00,0.00 & 0.11,0.11 & \textbf{\textless0.0001},\textbf{\textless0.0001}\\
Minor & 1664776380.0 & 0.00,0.00 & 0.01,0.01 & \textbf{\textless0.0001},\textbf{\textless0.0001}\\
Major & 1667877088.0 & 0.00,0.00 & 0.01,0.01 & \textbf{\textless0.0001},\textbf{\textless0.0001}\\
Critical & 1631274846.5 & 0.00,0.00 & 0.07,0.07 & \textbf{\textless0.0001},\textbf{\textless0.0001}\\
    \bottomrule
    \end{tabular}
\end{table}

\begin{table}[h]
    \caption{\rev{Detailed statistical tests results for metrics before the change is applied.
    Accompanies Table~\ref{tbl:stats_parent}.
    SHA is Shapiro-Wilk, MWU is Mann-Whitney U test, the number of samples for not perfective is 77,630, for perfective the number is 47,852.
    The number of samples for not corrective is 90,258 and for corrective 35,124.
    For both samples the median, Shapiro-Wilk test statistic and p-value are given comma separated.}}
    \centering
    \begin{tabular}{lrrrr}
    \multicolumn{5}{c}{Perfective changes}\\
    \toprule
Metric & MWU Statistic & Median & SHA Statistic & SHA p-val\\
    \midrule
McCC & 1946723702.0 & 47.00,39.00 & 0.27,0.21 & \textbf{\textless0.0001},\textbf{\textless0.0001}\\
LLOC & 1946637361.5 & 397.00,335.00 & 0.26,0.21 & \textbf{\textless0.0001},\textbf{\textless0.0001}\\
NLE & 1934702498.0 & 21.00,18.00 & 0.28,0.24 & \textbf{\textless0.0001},\textbf{\textless0.0001}\\
NUMPAR & 1860319211.0 & 34.00,32.00 & 0.25,0.20 & \textbf{\textless0.0001},\textbf{\textless0.0001}\\
CC & 1881849087.0 & 0.09,0.08 & 0.07,0.10 & \textbf{\textless0.0001},\textbf{\textless0.0001}\\
CLOC & 1642584608.5 & 84.00,118.00 & 0.18,0.24 & \textbf{\textless0.0001},\textbf{\textless0.0001}\\
CD & 1570226793.0 & 0.40,0.54 & 0.08,0.14 & \textbf{\textless0.0001},\textbf{\textless0.0001}\\
AD & 1548982847.0 & 0.83,1.00 & 0.11,0.14 & \textbf{\textless0.0001},\textbf{\textless0.0001}\\
NOA & 1861405551.0 & 2.00,2.00 & 0.13,0.09 & \textbf{\textless0.0001},\textbf{\textless0.0001}\\
CBO & 2023896520.5 & 21.00,15.00 & 0.24,0.16 & \textbf{\textless0.0001},\textbf{\textless0.0001}\\
NII & 1756171669.5 & 15.00,18.00 & 0.25,0.21 & \textbf{\textless0.0001},\textbf{\textless0.0001}\\
Minor & 1926916681.5 & 15.00,13.00 & 0.15,0.13 & \textbf{\textless0.0001},\textbf{\textless0.0001}\\
Major & 2025070328.5 & 4.00,3.00 & 0.23,0.17 & \textbf{\textless0.0001},\textbf{\textless0.0001}\\
Critical & 1949852017.5 & 0.00,0.00 & 0.19,0.12 & \textbf{\textless0.0001},\textbf{\textless0.0001}\\
    \midrule
    \multicolumn{5}{c}{Corrective changes}\\
    \midrule
McCC & 1455448657.0 & 41.00,50.00 & 0.23,0.23 & \textbf{\textless0.0001},\textbf{\textless0.0001}\\
LLOC & 1506999970.0 & 361.00,399.00 & 0.23,0.23 & \textbf{\textless0.0001},\textbf{\textless0.0001}\\
NLE & 1477296467.5 & 18.00,22.00 & 0.25,0.25 & \textbf{\textless0.0001},\textbf{\textless0.0001}\\
NUMPAR & 1573653093.5 & 33.00,33.00 & 0.22,0.22 & \textbf{\textless0.0001},\textbf{\textless0.0001}\\
CC & 1605078855.0 & 0.09,0.08 & 0.09,0.09 & \textbf{\textless0.0001},\textbf{\textless0.0001}\\
CLOC & 1683529860.5 & 101.00,83.00 & 0.22,0.22 & \textbf{\textless0.0001},\textbf{\textless0.0001}\\
CD & 1832629967.5 & 0.51,0.35 & 0.12,0.12 & \textbf{\textless0.0001},\textbf{\textless0.0001}\\
AD & 1822953682.5 & 1.00,0.75 & 0.13,0.13 & \textbf{\textless0.0001},\textbf{\textless0.0001}\\
NOA & 1616227506.0 & 2.00,2.00 & 0.10,0.10 & \textbf{\textless0.0001},\textbf{\textless0.0001}\\
CBO & 1476937730.0 & 17.00,21.00 & 0.19,0.19 & \textbf{\textless0.0001},\textbf{\textless0.0001}\\
NII & 1655048856.5 & 17.00,14.00 & 0.22,0.22 & \textbf{\textless0.0001},\textbf{\textless0.0001}\\
Minor & 1557520234.5 & 14.00,14.00 & 0.14,0.14 & \textbf{\textless0.0001},\textbf{\textless0.0001}\\
Major & 1516141644.5 & 3.00,4.00 & 0.19,0.19 & \textbf{\textless0.0001},\textbf{\textless0.0001}\\
Critical & 1546362144.0 & 0.00,0.00 & 0.15,0.15 & \textbf{\textless0.0001},\textbf{\textless0.0001}\\
    \bottomrule
    \end{tabular}
\end{table}

\end{appendices}

\end{document}